\documentclass[twocolumn, times, tighten]{aastex62}
\usepackage{textcomp}
\usepackage{amsmath}
\usepackage{color}
\usepackage{amssymb}
\usepackage{hyperref}

\newcommand{\FeII}{Fe\,{\sc ii}}
\newcommand{\Rfe}{$R_{\mathrm{Fe\,{II}}}$}

\newcommand{\OIII}{[O\,{\sc iii}]}
\newcommand{\Hbeta}{H$\beta$}

\shortauthors{Sun et al.}
\shorttitle{Quasar variability and its main sequence}

\begin{document}
\revised{\bf Draft: \today}
\title{Evolution of quasar stochastic variability along its main sequence}

\author[0000-0002-0771-2153]{Mouyuan Sun}
\affiliation{CAS Key Laboratory for Research in Galaxies and Cosmology, 
Department of Astronomy, University of Science and Technology of China, Hefei 
230026, China; ericsun@ustc.edu.cn; xuey@ustc.edu.cn}
\affiliation{School of Astronomy and Space Science, University of Science 
and Technology of China, Hefei 230026, China}

\author[0000-0002-1935-8104]{Yongquan Xue}
\affiliation{CAS Key Laboratory for Research in Galaxies and Cosmology, 
Department of Astronomy, University of Science and Technology of China, Hefei 
230026, China; ericsun@ustc.edu.cn; xuey@ustc.edu.cn}
\affiliation{School of Astronomy and Space Science, University of Science 
and Technology of China, Hefei 230026, China}

\author[0000-0002-4419-6434]{Junxian Wang}
\affiliation{CAS Key Laboratory for Research in Galaxies and Cosmology, 
Department of Astronomy, University of Science and Technology of China, Hefei 
230026, China; ericsun@ustc.edu.cn; xuey@ustc.edu.cn}
\affiliation{School of Astronomy and Space Science, University of Science 
and Technology of China, Hefei 230026, China}

\author[0000-0002-4223-2198]{Zhenyi Cai}
\affiliation{CAS Key Laboratory for Research in Galaxies and Cosmology, 
Department of Astronomy, University of Science and Technology of China, Hefei 
230026, China; ericsun@ustc.edu.cn; xuey@ustc.edu.cn}
\affiliation{School of Astronomy and Space Science, University of Science 
and Technology of China, Hefei 230026, China}

\author[0000-0001-8416-7059]{Hengxiao Guo}
\affiliation{National Center for Supercomputing Applications, University of 
Illinois at Urbana-Champaign, 605 East Springfield Avenue, Champaign, IL 61820, 
USA}
\affiliation{Department of Astronomy, University of Illinois at Urbana-Champaign, 
Urbana, IL 61801, USA}

\begin{abstract}
We explore the evolution of the time variability (in the optical 
$g$-band and on timescales of weeks to years) of SDSS Stripe 82 quasars 
along the quasar main sequence. A parent sample of $1004$ quasars 
within $0.5\leq z \leq 0.89$ are used for our statistical studies; we then 
make subsamples from our parent sample: a subsample of $246$ quasars with similar 
luminosities, and a subsample of $399$ quasars with similar \Rfe\ (i.e., the ratio 
of the equivalent width of \FeII\ within $4435$--$4685\ \mathrm{\AA}$ to that of 
\Hbeta). We find the variability amplitude decreases with luminosity 
($L_{\mathrm{bol}}$). The anti-correlation between the variability amplitude and 
\Rfe\ is weak but statistically significant. The characteristic timescale, $\tau$, 
correlates mostly with quasar luminosity; its dependence on \Rfe\ is 
statistically insignificant. After controlling luminosity and \Rfe, the 
high- and low-FWHM samples have similar structure functions. These results 
support the framework that \Rfe\ is governed by Eddington ratio and FWHM 
of \Hbeta\ is mostly determined by orientation. We then provide 
new empirical relations 
between variability parameters and quasar properties (i.e., luminosity and \Rfe). 
Our new relations are consistent with the scenario that quasar variability is 
driven by the thermal fluctuations in the accretion disk; $\tau$ seems to 
correspond to the thermal timescale. From our new relations, we find the 
short-term variability is mostly sensitive to $L_{\mathrm{bol}}$. Basing on 
this, we propose that quasar short-term (a few months) variability might be 
a new type of ``Standard Candle'' and can be adopted to probe cosmology. 
\end{abstract}


\section{Introduction}
\label{sect:intro}
Quasars\footnote{We use the term “quasar” to generically refer to active galactic 
nuclei (AGNs) with optical broad emission lines, regardless of luminosity.} show 
aperiodic luminosity variations across the electromagnetic spectrum \citep[for 
a review, see][]{Ulrich1997}. The physical nature of quasar variability remains 
unclear although a number of theoretical scenarios have been proposed. For 
instance, the local \citep{Lyubarskii1997} or the global accretion rate 
\citep{Li2008} fluctuations can induce variations in quasar luminosity and have 
the potential to explain the power spectral density (PSD) and the amplitude of 
quasar variability. It is also speculated that quasar variability is driven by 
the thermal fluctuations in accretion disk \citep[e.g.,][]{Czerny1999, Kelly2013}. 
Moreover, the ultraviolet (UV) or optical variations on short timescales might 
also be induced by X-ray reprocessing \citep{Czerny1999, Kubota2018}. X-ray 
reprocessing could also be responsible for the inter-band time lags \citep[][but 
see \citealt{Shappee2014, Fausnaugh2016, Starkey2016, Gardner2017, 
Starkey2017, Zhu2017}]{Krolik1991, Edelson1996, Edelson2015, Edelson2017, 
Wanders1997, Collier1998, Sergeev2005, McHardy2014, McHardy2016, Cackett2017, 
McHardy2017, Sun2018}. 

Different physical scenarios manifest as various correlations between the variability 
parameters and quasar properties. Indeed, previous works on both individual 
and ensemble quasar variability have revealed that the amplitude and the PSD 
shape depend on quasar luminosity ($L_{\rm{bol}}$), the mass ($M_{\rm{BH}}$) 
of the supermassive black hole (SMBH), and wavelength \citep[see 
e.g.,][]{Uomoto1976, Hook1994, Giveon1999, Hawkins2002, Berk2004, Vries2005, 
Wilhite2008, Bauer2009, Kelly2009, Macleod2010, Macleod2012, Zuo2012, Kelly2013, 
Sun2015, Kozlowshi2016, Guo2017}. Roughly speaking, these correlations are not 
entirely consistent with theoretical expectations. For instance, according to 
the classical thin disk theory \citep{SSD}, the thermal timescale 
($\tau_{\rm{TH}}$) for a fixed wavelength depends only on quasar bolometric 
luminosity, i.e., $\tau_{\rm{TH}}\propto L_{\rm{bol}}^{1/2}$. However, 
\cite{Macleod2010} constrained the characteristic timescale ($\tau$) of quasar 
variability by fitting the \textit{continuous time first-order autoregressive 
process} \citep[i.e., CAR(1), whose PSD has the following shape $\mathrm{PSD}(f) 
\propto 1/(f_0^2+f^2)$, where $f_0=1/\tau$; see, e.g.,][and 
Section~\ref{sect:definition}]{Kelly2009, Kozlowski2010} to the light curves 
of the Sloan Digital Sky Survey (SDSS) Stripe 82 (S82) quasars and investigated 
the scaling relation between $\tau$ and $L_{\rm{bol}}$ and $M_{\rm{BH}}$; they 
found that the best-fitting scaling relation is incompatible with the expected 
scaling relations for the thermal or the viscous timescales. It is unclear 
whether the discrepancy is real or is simply caused by some systematic biases 
in estimating the variability parameters and quasar properties. 

\cite{Macleod2012} and \cite{Guo2017} argued that the PSD of the observed light 
curves on long timescales (i.e., $\gg \tau$) should be steeper than that of 
the CAR(1) process. The deviation from the CAR(1) process on short timescales 
(i.e., sub-month) has also been proposed \citep[e.g.,][]{Mushotzky2011, Kasliwal2015, 
Simm2016, Caplar2017, Smith2018}. 

Recently, \cite{Kozlowshi2017} explored the biases of the estimation of 
$\tau$ via fitting the CAR(1) process to individual light curves. They concluded 
that $\tau$ and other variability parameters are incorrectly determined if 
the baseline is too short, and the reported scaling relations between the 
variability parameters and quasar properties are unlikely to be robust. 
Instead, the ensemble structure function (which measures the variability 
amplitude as a function of timescale; see Section~\ref{sect:definition}) 
is found to be less biased \citep{Kozlowshi2016}.  

$M_{\rm{BH}}$, one of the key parameters of SMBH, is hard to be robustly 
measured for quasars. The most widely adopted approach is via the single-epoch 
virial black hole mass estimators (e.g., \citealt{Vestergaard2002, Vestergaard2006, 
Shen2011}; for a recent review, see \citealt{Shen2013}). These estimators 
are based on two assumptions: first, the broad emission line region (BLR) 
radius-quasar luminosity relation is valid for the full quasar population; 
second, the line widths of the broad emission lines (BELs) trace the virial 
motions of the BLR gas. The empirical BLR radius-quasar luminosity relation 
\citep[e.g.,][]{Bentz2013} is derived from a small sample of 
sources.\footnote{The on-going SDSS-RM program can greatly enlarge the sample 
size \citep[e.g.,][]{Shen2015, Grier2017b}.} There is new evidence that this 
empirical relation is invalid for high Eddington ratio sources \citep{Du2014}. 

Quasar spectra show diverse features in terms of emission lines. It is shown 
that the diversity can be well represented by several eigenvectors. It is 
widely speculated that the Eigenvector 1 (hereafter, EV1), which is the main 
variance of the diversity, is driven by Eddington ratio \citep{Boroson1992, 
Sulentic2000a, Sulentic2000b, Boroson2002, Runnoe2014}. \cite{Shen2014} and 
\cite{SunJY2015} adopted the orientation independent $M_{\mathrm{BH}}$ indicators 
and found that, after controlling for quasar luminosity, the \FeII\ strength, 
\Rfe\ (i.e., the ratio of the equivalent width of \FeII\ within $4435$--$4685\ 
\mathrm{\AA}$ to that of \Hbeta), anti-correlates with $M_{\mathrm{BH}}$; after 
controlling quasar luminosity and \Rfe, the correlation between FWHM and 
$M_{\mathrm{BH}}$ is rather weak or absent; it is likely that the line widths 
of BELs are sensitive to inclination \citep[see also][]{Collin2006, 
Runnoe2013, Pancoast2014, Bisogni2017, Grier2017a, Storchi2017}. 
Therefore, quasars can be unified by Eddington ratio (or \Rfe\ ) and orientation 
(or the line width of \Hbeta) \citep[i.e., the quasar main sequence; see 
e.g.,][]{Shen2014}. 

It is interesting to investigate the evolution of quasar variability on the 
main-sequence plane. There are only a few studies of this topic. For instance, 
\cite{Ai2010} focused on the tight correlation between the long-term variability 
amplitude\footnote{It is well known that the PSD of quasar variability increases 
with timescales. Therefore, the excess of variance of a long light curve reflects 
the long-term variability.} and \Rfe . 

In order to better understand the relationship between quasar variability and 
the main sequence, and to test the physical scenarios, we study the $g$-band 
light curves of spectroscopically confirmed SDSS S82 quasars by calculating 
the ensemble structure functions along the main sequence. We choose 
$g$-band for two reasons: first, compared with $r$-band, $g$-band is less 
contaminated by galaxy emission; second, the noise level of $g$-band is smaller 
than that of $u$-band. 

This paper is formatted as follows. In Section~\ref{sect:sample}, we 
introduce our sample selection. In Section~\ref{sect:definition}, we 
describe the structure function and the CAR(1) process. In 
Section~\ref{sect:sf_phys}, we present quasar variability along the main 
sequence. In Section~\ref{sect:disall}, we discuss the implications of 
our results. We summarize our main conclusions in Section~\ref{sect:summ}. 
In this work, we adopt a flat $\Lambda$CDM cosmology with $h_0=0.7$ 
and $\Omega_{\mathrm{m}}=0.3$ unless otherwise specified.

\section{Sample selection}
\label{sect:sample}
Our initial parent sample consists of the SDSS S82 quasars considered by \cite{Macleod2010}. 
The S82 quasars have on average $\sim 60$ epochs of accurate photometry in five 
bands \citep[i.e., $ugriz$; see][]{Gunn2006}; these light curves 
can effectively probe rest-frame timescales from weeks to six years. The light curve 
data\footnote{The data set can be accessed from 
\url{http://faculty.washington.edu/ivezic/macleod/qso_dr7/Southern.html}.} are 
produced with improved calibration techniques \citep{Ivezic2007, Sesar2007}. We 
then cross match this parent sample with the catalog of quasar properties from 
SDSS DR7 \citep{Shen2011} and obtain the emission line properties and quasar 
parameters (e.g., the bolometric luminosity, $L_{\rm{bol}}$). As a second step, 
we only select quasars with available properties of \Hbeta\ and Fe within $4435\ 
\rm{\AA}$--$4685\ \rm{\AA}$. Radio-loud (i.e., radio loudness $R= 
f_{\nu}(6\ \mathrm{cm})/f_{\nu}(2500\ \mathrm{\AA})>10$) quasars are 
also rejected. The resulting parent sample that will be used for our subsequent 
studies has $1004$ quasars within $0.5\leq z \leq 0.89$. We only consider sources 
in such a narrow range of redshift to eliminate the rest-frame wavelength dependence. 

\begin{figure}
\epsscale{1.2}
\plotone{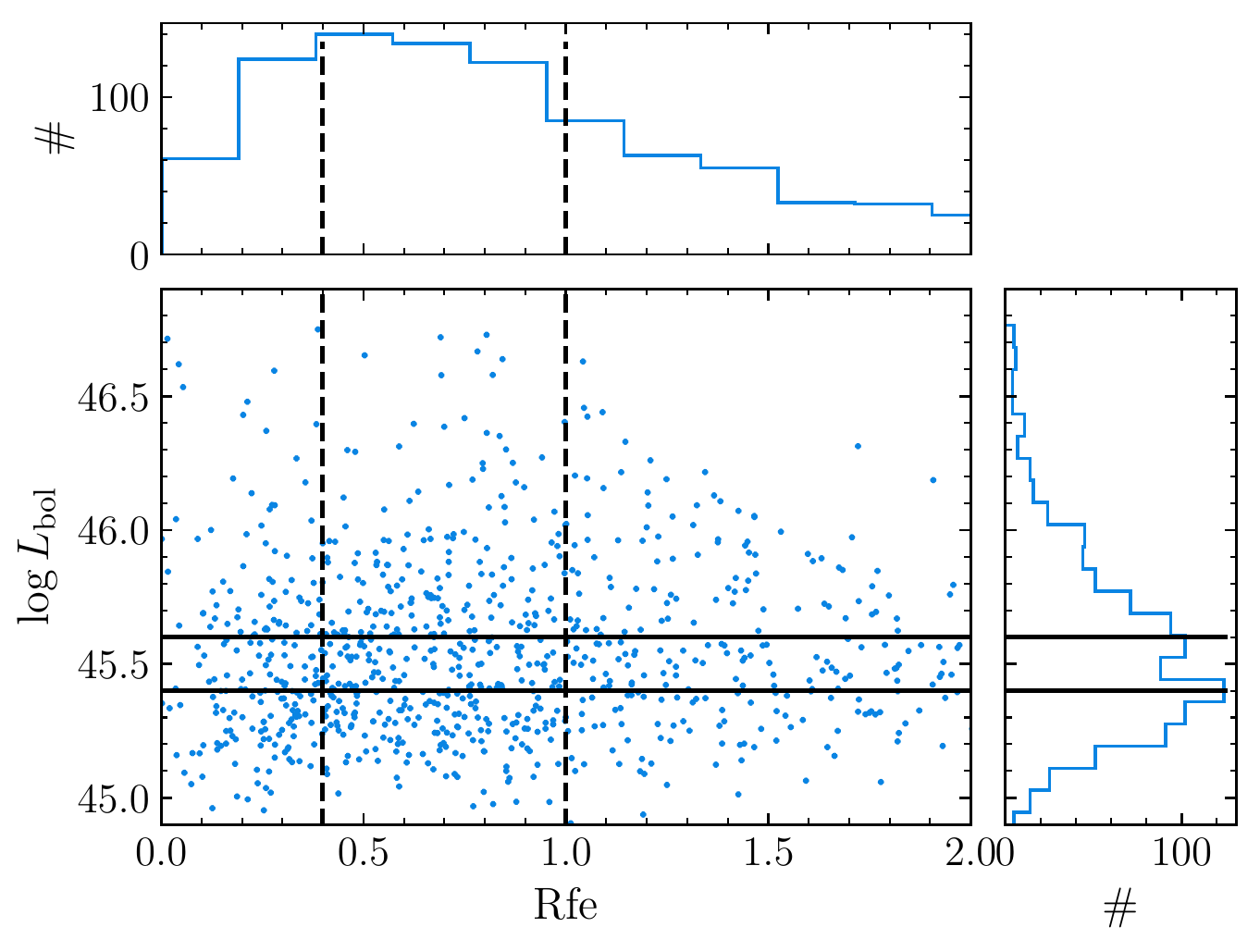}
\caption{Distribution of our parent sample in the \Rfe-$L_{\rm bol}$ plane. The 
two vertical dashed lines define the \Rfe-matched sample. The two horizontal solid 
lines indicate the $L_{\rm bol}$-matched sample. }
\label{fig:f1}
\end{figure}

The distribution of our parent sample in the \Rfe-$L_{\rm bol}$ plane is shown 
in Figure~\ref{fig:f1}. To explore the relationship between quasar variability 
and the main sequence, we make subsamples from our parent sample: a subsample 
of quasars with similar luminosity and redshift (i.e., the luminosity-matched 
sample), and a subsample of quasars with similar \Rfe\ and redshift (i.e., the 
\Rfe-matched sample). 

\begin{deluxetable*}{cccccc}
\tablecaption{Properties of the luminosity- and \Rfe- matched samples \label{table:prop}}
\tablehead{\colhead{} & \colhead{} & \colhead{Number} & \colhead{$\log\ 
L_{\mathrm{bol}}$} & \colhead{FWHM\ (\Hbeta)} & \colhead{\Rfe} \\ \colhead{} & 
\colhead{} & \colhead{} & \colhead{($\mathrm{erg\ s^{-1}}$)} & 
\colhead{($\mathrm{km\ s^{-1}}$)} & \colhead{} } 
\startdata
{} & High-\Rfe\ bin & 82 &  45.50$\pm$0.01 &  3170$\pm$220 &  1.83$\pm$0.07 \\
{The $L_{\rm bol}$-matched sample} & Middle-\Rfe\ bin & 82 &  45.50$\pm$0.01 &  
3410$\pm$260 &  0.89$\pm$0.03 \\
{} & Low-\Rfe\ bin & 82 &  45.49$\pm$0.02 & 4590$\pm$300 &  0.40$\pm$0.02 \\
\hline
{} & High-$L_{\mathrm{bol}}$ bin & 132 &  45.87$\pm$0.02 &  4440$\pm$220 &  0.70$\pm$0.02 \\
{The \Rfe-matched sample} & Middle-$L_{\mathrm{bol}}$ bin & 134 &  45.50$\pm$0.01 &  
4210$\pm$320 &  0.66$\pm$0.04 \\
{} & Low-$L_{\mathrm{bol}}$ bin & 133 &  45.25$\pm$0.01 &  4550$\pm$330 &  0.70$\pm$0.02 \\
\enddata
\tablecomments{The quoted value is the median of each bin. The $1\sigma$ error bar 
is calculated via bootstrapping. }
\end{deluxetable*}

The luminosity-matched sample: this sample initially consists of $246$ quasars 
with $10^{45.4} \ \mathrm{erg\ s^{-1}}\leq L_{\mathrm{bol}} \leq 10^{45.6}\ 
\mathrm{erg\ s^{-1}}$ and $0.5\leq z \leq 0.89$ (i.e., the region defined by 
two solid lines in Figure~\ref{fig:f1}). We choose such a narrow luminosity 
range for two reasons. First, the distribution of $L_{\rm bol}$ peaks at this 
luminosity range (see Figure~\ref{fig:f1}). Second, the variability amplitude 
depends critically upon $L_{\rm bol}$ but weakly on \Rfe\ (see 
Section~\ref{sect:sf_rfe}). We have verified that our conclusions would not change 
if we, for instance, consider quasars in other luminosity ranges. This sample 
is further divided into three bins in \Rfe, with each having one third of quasars. 
To ensure that quasars in the three bins have similar distributions of $L_{\rm{bol}}$, 
we apply the Anderson-Darling test to the three bins. The null hypothesis of 
this test is that quasars in the three bins are drawn from the same population 
of $L_{\mathrm{bol}}$. If the null hypothesis is rejected (i.e., the $p$-value 
$\leq 0.05$), for each bin, we clip the $1$D distributions of $L_{\rm{bol}}$ 
so that only objects within $1$th--$99$th percentiles are included. 
Then, the Anderson-Darling test is applied to the new three bins. We repeat 
this process until the null hypothesis cannot be rejected (i.e., the $p$-value 
$>0.05$). During this process, no source is discarded because of the narrow 
luminosity range. The properties of the three bins are summarized in 
Table~\ref{table:prop}. 

The \Rfe-matched sample: this sample initially consists of $412$ quasars with 
$0.4\leq$ \Rfe $\leq 1.0$ and $0.5\leq z \leq 0.89$ (i.e., the region 
defined by two dashed lines in Figure~\ref{fig:f1}). This sample is also further 
divided into three bins in $L_{\rm{bol}}$, with each having one third of quasars. 
Similar to that of the luminosity-matched sample, we use the same approach to 
ensure the three bins are consistent with being drawn from the same population 
of \Rfe. During this process, $13$ sources are rejected. The properties of the 
three bins are summarized in Table~\ref{table:prop}.

\section{Definition of structure function and the CAR(1) process}
\label{sect:definition}
\subsection{Structure function}
The structure function\footnote{For a general discussion, see, e.g., 
\cite{Emmanoulopoulos2010}, \cite{Kozlowshi2016}.}, $\mathrm{SF}(\Delta t)$, measures 
the statistical dispersion of two random variables (i.e., a magnitude pair) separated 
by time intervals, $\Delta t$. The structure function can be used to characterize the 
statistical dispersion of $\Delta m$ for a sample of many similar quasars with the 
same (or close) $\Delta t$, where $\Delta m$ is the magnitude difference between two 
observations. We adopted the interquartile range (i.e., IQR) to measure the statistical 
dispersion as it is robust against outliers or tails in the distribution. Therefore, 
we calculate the statistical dispersion as follows \citep{Macleod2010, Sun2015}, 
\begin{equation}
\label{eq:iqr}
\mathrm{SF_{IQR}}(\Delta t) = 0.74\mathrm{IQR}(\Delta m) \\,
\end{equation}
where $\mathrm{IQR}(\Delta m)$ is the $25\%-75\%$ interquartile range of $\Delta m$. 
The constant $0.74$ normalizes the IQR to be equivalent to the standard deviation of 
a Gaussian distribution. Therefore, $0.74 \mathrm{IQR}$ is known as the normalized IQR 
(hereafter NIQR). 

It should be noted that the measured statistical dispersion (i.e., Eq.~\ref{eq:iqr}) 
is a superposition of measurement errors and quasar variability. On very short timescales 
(e.g., days), the amplitude of quasar variability is small and the statistical dispersion 
is dominated by measurement errors. Therefore, we can estimate measurement errors 
from the statistical dispersion on timescales of a few days. On timescales of months to 
years, the contribution of measurement errors becomes negligible.

\subsection{The CAR(1) process}
The CAR(1) process is often referred as the damped random walk (DRW) or the Ornstein-Uhlenbeck 
(OU) process; this process is proven to be effective in describing the light curves of 
quasar continuum emission \citep[e.g.,][]{Kelly2009, Kozlowski2010, Macleod2010, Macleod2012, 
Zu2013}. The structure function of the CAR(1) process is given by 
\begin{equation}
\label{eq:sfdrw}
\mathrm{SF}(\Delta t|\tau, \hat{\sigma})=\hat{\sigma}\sqrt{\tau(1-\exp(-\Delta t/\tau))} \\,
\end{equation}
where $\Delta=|t_i-t_j|$ is the separation time between two observations. That is, the CAR(1) 
process is characterized by two parameters, $\hat{\sigma}$ and $\tau$. $\hat{\sigma}$ 
determines the short-term variability amplitude; $\tau$ is the characteristic timescale. 

It should be noted that quasar variability might be more complex than the CAR(1) 
process. Therefore, \cite{Kelly2014} proposed more flexible continuous-time autoregressive 
moving average (i.e., CARMA($p$,$q$)) models to describe quasar light curves; the CAR(1) process 
corresponds to the CARMA(1,0) process. For each source in our parent sample, we use the Python 
CARMA package\footnote{This package can be downloaded from 
\url{https://github.com/brandonckelly/carma_pack}.} and adopt the Akaike information criterion 
\citep[AIC;][]{Akaike1974} to choose the order of the CARMA($p$,$q$) models (i.e., determining 
$p$ and $q$ that minimize AIC; see section 3.5 of \citealt{Kelly2014}); we also calculated 
AIC for the CAR(1) process (hereafter AIC(1)). We found that, for most of our light curves ($\sim 
90\%$), the differences between the minimum AIC and AIC(1) is less than $10$. Therefore, it 
seems that the data quality of our sample is insufficient to distinguish between the CAR(1) process and 
other more complex models. In Section~\ref{sect:dis2}, we will model the structure functions 
as the CAR(1) process (i.e., Eq.~\ref{eq:sfdrw}); however, more complex models (i.e., 
Eq.~\ref{eq:sfmodel}) are also discussed. If quasar variability is indeed not driven by the 
stochastic models we assumed or the light curve is a nonstationary process, the uncertainties 
of our model parameters in Section~\ref{sect:dis2} and Tables~\ref{table:link_fit} and \ref{table:fitfun} 
might be inaccurate \citep[or even underestimated; see e.g.,][]{White1982}.

\section{The Ensemble Structure Function and Quasar Main Sequence}
\label{sect:sf_phys}
\subsection{The Ensemble Structure Function and \Rfe}
\label{sect:sf_rfe}
\begin{figure}
\epsscale{1.2}
\plotone{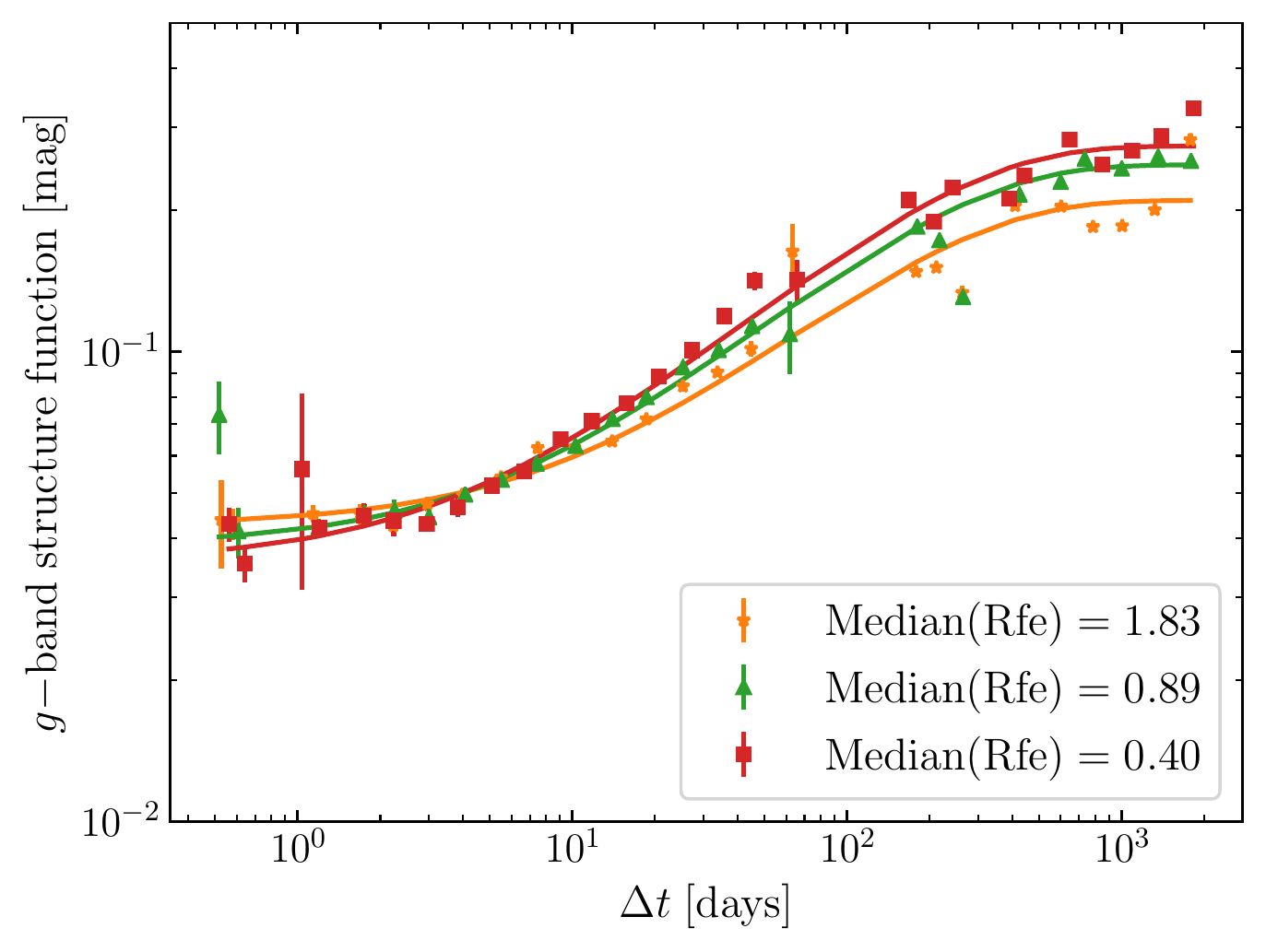}
\caption{The $g$-band ensemble structure functions for the three bins, controlling 
$L_{\rm bol}$ and $z$. Low-\Rfe\ quasars are more variable. The solid lines represent 
our best-fitting models (see Section~\ref{sect:dis2}).}
\label{fig:sf_rfe}
\end{figure}

We aim to explore the ensemble variability of quasar continuum as a function of Rfe. The 
ensemble structure functions for the three bins of the luminosity-matched sample are presented 
in Figure~\ref{fig:sf_rfe}. Low-\Rfe\ quasars tend to be more variable (for 
a statistical description of our conclusion, see Section~\ref{sect:dis2}). Our result is well expected 
if: (1) EV1 is indeed driven by Eddington ratio and (2) for fixed luminosity, high Eddington 
ratio quasars are more stable. The former assumption is supported by independent tests 
\citep[e.g.,][]{Shen2014, SunJY2015}. We will discuss possible explanations of the second 
requirement in Section~\ref{sect:dis3}. 

\begin{figure}
\epsscale{1.2}
\plotone{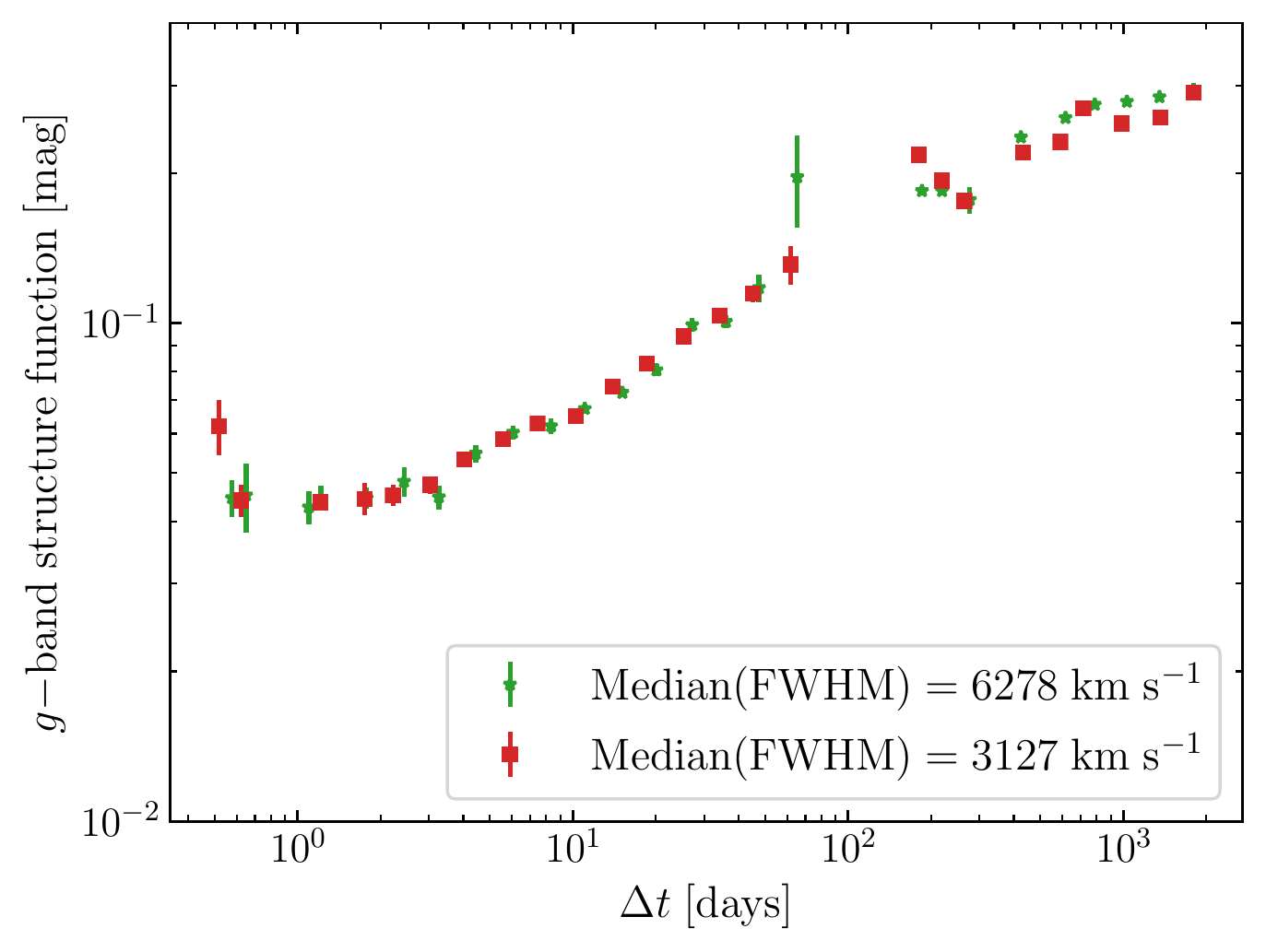}
\caption{The $g$-band ensemble structure functions for the high- and low-FWHM bins, 
controlling bolometric luminosity, redshift and \FeII\ strength. The two samples have similar 
structure functions. Hence, quasar variability and FWHM are intrinsically uncorrelated or the 
correlation is rather weak. }
\label{fig:sf_fwhm}
\end{figure}

The tendency between \Rfe\ and the quasar variability amplitude might be 
induced by FWHM of \Hbeta\footnote{Throughout this work, FWHM refers to \Hbeta , 
unless otherwise specified.} since there might be an anti-correlation between FWHM and 
\Rfe\ (see Table~\ref{table:prop}) and $M_{\mathrm{BH}}\propto \mathrm{FWHM}^2$. In 
order to verify this speculation, we explore quasar variability as a function of FWHM after controlling 
\Rfe, $L_{\rm bol}$ and $z$. Therefore, we construct samples as follows. First, we select 
quasars within $10^{45.3} \ \mathrm{erg\ s^{-1}}\leq L_{\mathrm{bol}} \leq 10^{45.6}\ 
\mathrm{erg\ s^{-1}}$ and $0.4<$\Rfe $<1$. We now choose a slightly wider luminosity bin to 
increase the statistic. Second, these sources are divided into two bins according 
to FWHM, i.e., the low- (high-) FWHM bin with $\mathrm{FWHM}$ being smaller (larger) than 
$\mathrm{Median(FWHM)}$. Third, we ensure $L_{\rm bol}$ and \FeII\ strength of the two samples 
are matched via the methodology in Section~\ref{sect:sample}. The number of quasars in the 
low- (high-) FWHM bin is $75$ ($74$). The median values of FWHM for 
the two sub-samples are $3127\ \mathrm{km\ s^{-1}}$ and $6278\ \mathrm{km\ s^{-1}}$, 
respectively. As shown in Figure~\ref{fig:sf_fwhm}, the ensemble structure functions for 
the two sub-samples are quite similar. Therefore, it seems that the relation between quasar 
variability and the virial $M_{\rm BH}$ is rather weak or absent since, 
for fixed quasar luminosity, $M_{\mathrm{BH}}\propto \mathrm{FWHM}^2$. 

\begin{figure}
\epsscale{1.2}
\plotone{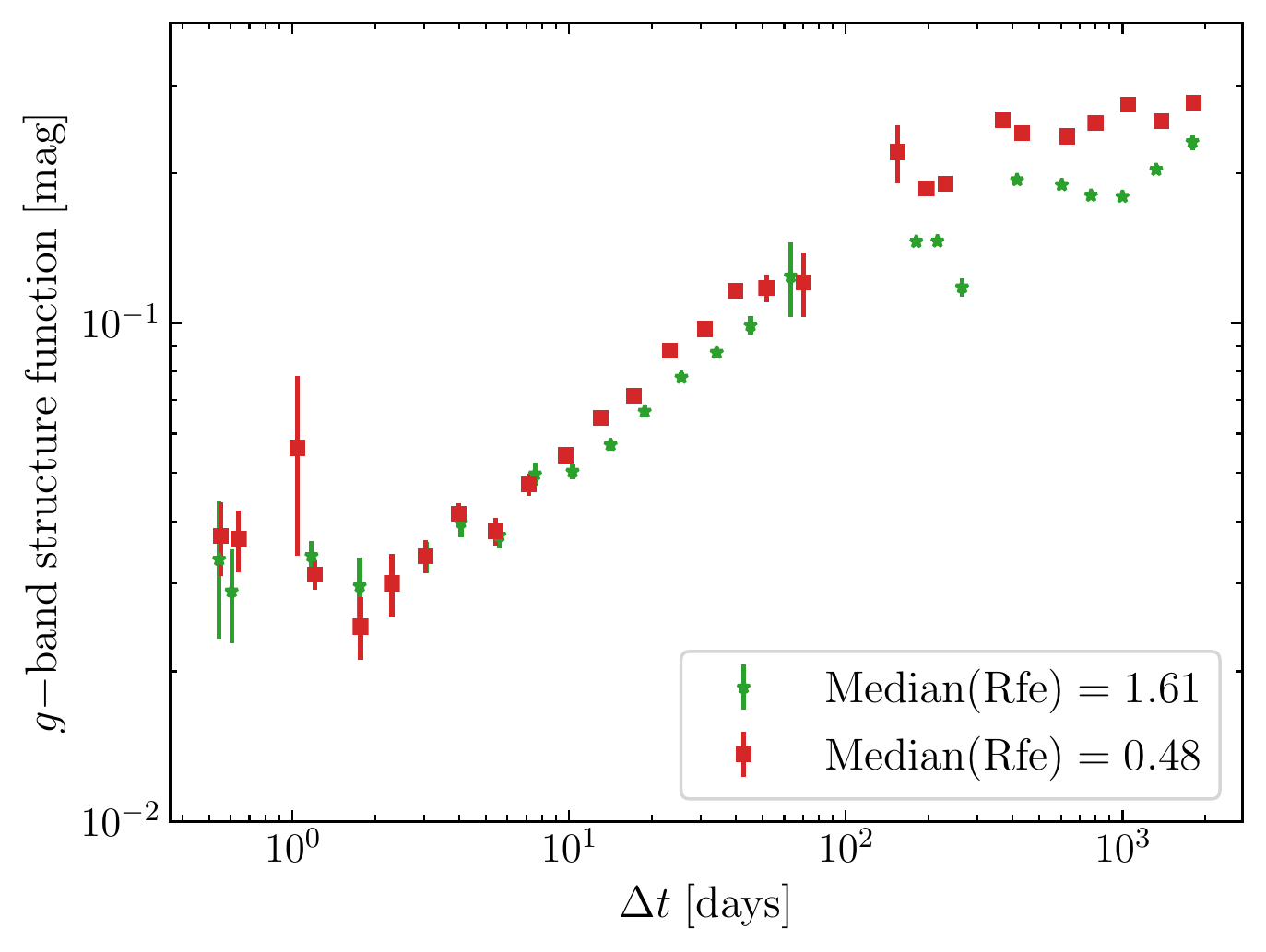}
\caption{The $g$-band ensemble structure functions for the large- and small-\Rfe\ bins, 
controlling bolometric luminosity, redshift and FWHM. Sources with larger \Rfe\ tend to 
be less variable. }
\label{fig:sf_rfe_fwhm}
\end{figure}

We also control FWHM, $L_{\rm bol}$, $z$, and divide sources into two \Rfe\ bins following the 
method we mentioned above. That is, we select quasars within $10^{45.3} \ \mathrm{erg\ s^{-1}} 
\leq L_{\mathrm{bol}} \leq 10^{45.6}\ \mathrm{erg\ s^{-1}}$ and $3000\ \mathrm{km\ s^{-1}}< 
\mathrm{FWHM}<5000\ \mathrm{km\ s^{-1}}$ and divide them into two bins according to \Rfe . We 
calculate the structure functions for the two bins. We again find that sources with larger \Rfe\ 
tend to be less variable (see Figure~\ref{fig:sf_rfe_fwhm}). These conclusions provide additional 
evidence supporting the claim that orientation determines the dispersion of FWHM 
\citep[e.g.,][]{Shen2014}. We will discuss this idea in Section~\ref{sect:dis1}. 

\subsection{The Ensemble Structure Function and Quasar Luminosity}
\label{sect:sf_lbol}
\begin{figure}
\epsscale{1.2}
\plotone{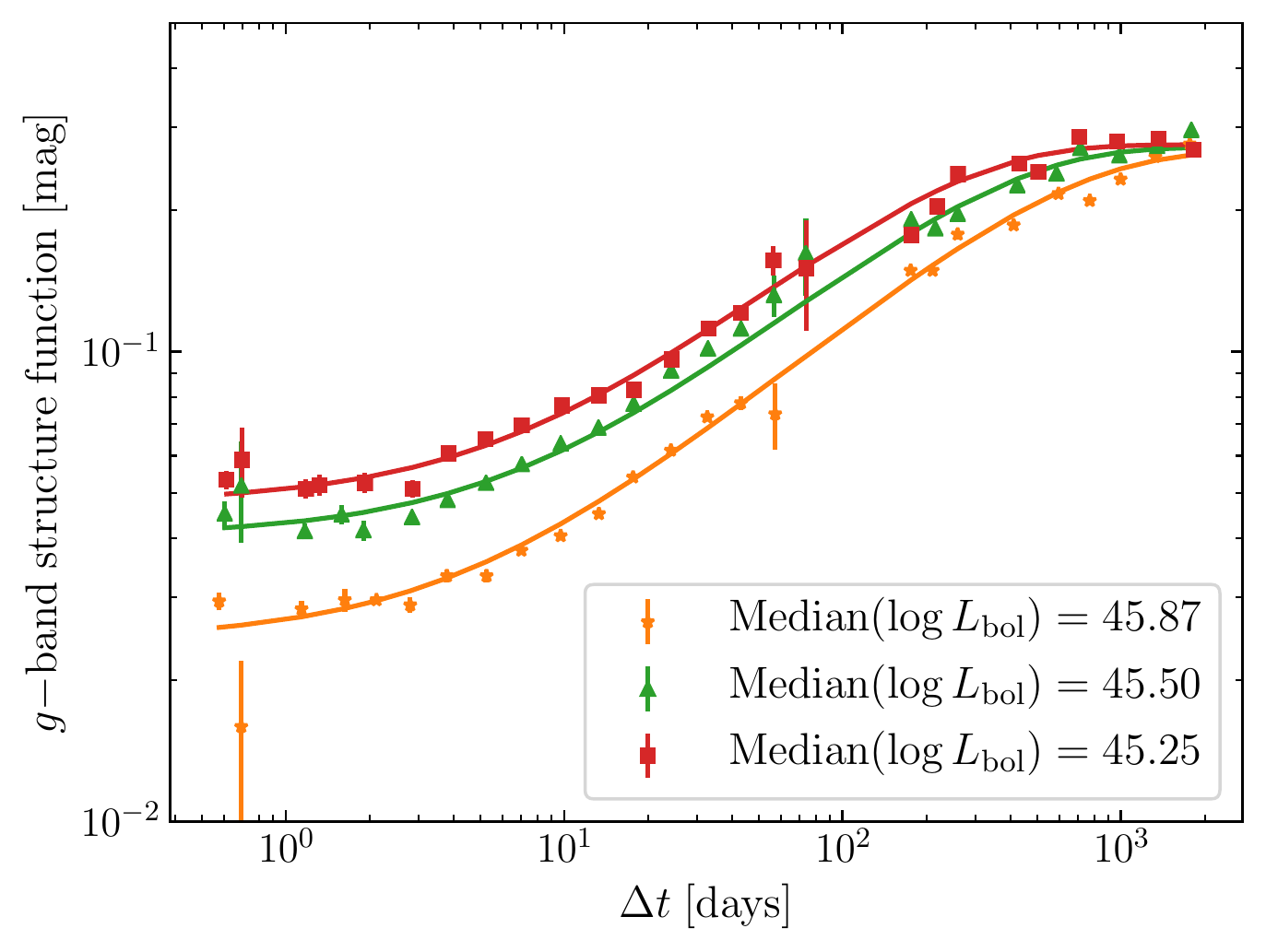}
\caption{The $g$-band ensemble structure functions for the three $L_{\mathrm{bol}}$ bins, 
controlling \Rfe\ and redshift. On short timescales (i.e., $1\la \Delta t\la 100$ days), quasar 
variability and $L_{\rm bol}$ are anti-correlated. This tendency diminishes on long timescales 
(i.e., $\Delta t\ga 100$days). The solid lines represent our best-fitting models (see 
Section~\ref{sect:dis2}). }
\label{fig:sf_lbol}
\end{figure}

In the previous section, we demonstrate the relation between quasar variability and \Rfe. 
To examine whether there is an additional dependence on $L_{\rm bol}$, we compare the 
ensemble structure functions of the \Rfe-matched sample (see Figure~\ref{fig:sf_lbol}). 
On short timescales (i.e., $1 \la \Delta t\la 100$ days), there is a clear anti-correlation 
between quasar variability and $L_{\rm bol}$ (for a statistical description 
of our conclusion, see Section~\ref{sect:dis2}). This tendency diminishes on long 
timescales (i.e., $\Delta t\ga 100$ days). Therefore, it seems that: (1) $L_{\rm bol}$ 
controls the short-term ($1 \la \Delta t\la 100$ days) quasar variability 
and (2) \Rfe\ drives quasar variability on timescales of $\Delta t \ga 10$ days. 

\section{Discussion}
\label{sect:disall}
\subsection{Implications to the Structure of BLR}
\label{sect:dis1}
According to our inspection of the structure function described in 
Section~\ref{sect:sf_phys}, quasar variability at a given wavelength in 
the UV/optical bands and on timescales from weeks to years can 
be characterized by $L_{\rm bol}$ and \Rfe. There is no additional correlation between quasar 
variability and FWHM. Our results can be well explained in the framework that the Eddington 
ratio and orientation govern most of the quasar diversity \citep{Shen2014}. According to this 
scenario, the EV1 is driven by the Eddington ratio; high \FeII\ strength sources have high 
Eddington ratios and are less variable; FWHM is a tracer of orientation and does not correlate 
with quasar variability. 

\begin{figure}
\epsscale{1.2}
\plotone{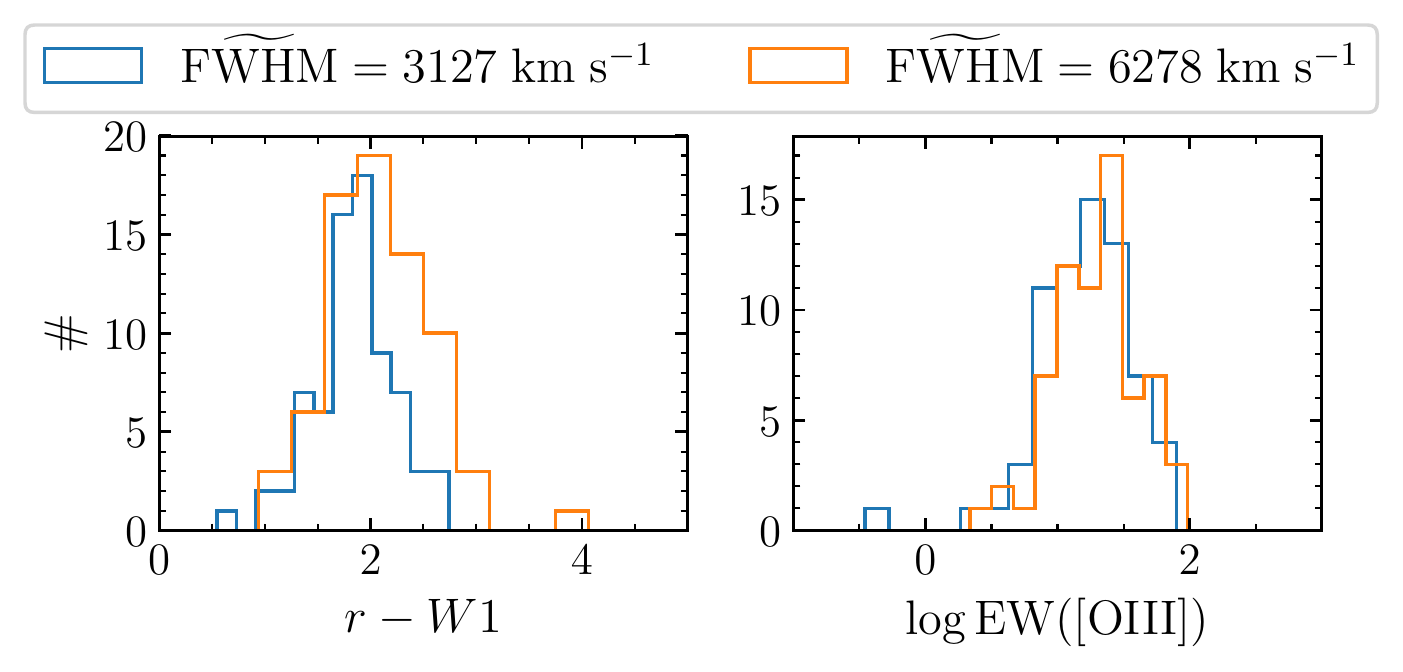}
\caption{Left: The distributions of $r-W1$ color for the broad- and narrow-FWHM bins, 
controlling $L_{\mathrm{bol}}$, redshift and \Rfe. Souces in the broad-FWHM bin tend 
to have redder $r-W1$ colors. Right: The distributions of EW(\OIII) for the broad- 
and narrow-FWHM bins, controlling $L_{\mathrm{bol}}$, redshift and \Rfe. The two bins 
are consistent with being drawn from the same population of EW(\OIII). }
\label{fig:cl_o3}
\end{figure}

To test whether FWHM traces orientation, we compare the $r-W1$ color of the low-FWHM sample 
with that of the high-FWHM sample, where $W1$ refers to the \textit{WISE} $3.4\ \mathrm{\mu m}$ 
band. To obtain $W1$, we cross-match our quasars with the ALLWISE 
catalog\footnote{The catalog is availabel at \url{http://wise2.ipac.caltech.edu/docs/release/allwise/}} 
\citep{Wright2010, Mainzer2011} with the 
maximum matching radius of $2^{''}$. The left panel of Figures~\ref{fig:cl_o3} presents our 
results. Indeed, sources in the broad-FWHM bin tend to have redder SED than the narrow-FWHM 
sample (the $p$ value of the Anderson-Darling test is $<0.01$). Similar results have been 
obtained by \cite{Shen2014}. Therefore, broad- (narrow-) FWHM sources are consistent with 
being viewed more edge-(face-) on. If so, the geometry of BLR is disk-like rather than 
spherical, which is consistent with other observations \citep[e.g.,][]{Jarvis2006, 
Pancoast2014, Grier2017a, Storchi2017,Xiao2018}. 
The orientation scenario also naturally explains the lack of correlation between 
quasar variability and FWHM (Figure~\ref{fig:sf_fwhm}).  

\OIII\ EW has also been proposed as a tracer of orientation \citep[e.g.,][]{Risaliti2011}. 
We also show the distributions of \OIII\ EW for the broad and narrow FWHM samples in the 
right panel of Figure~\ref{fig:cl_o3}. Contrary to our expectation, we cannot reject the 
null hypothesis that the two distributions of \OIII\ EW are drawn from the same population 
(the $p$ value of the Anderson-Darling test is $0.4$). Therefore, we conclude that \OIII\ 
EW is driven by \Rfe\ \citep[i.e., the EV1; ][]{Boroson1992} or the maximum disk temperature 
\citep{Panda2017} rather than orientation.

\subsection{Modeling Quasar Variability}
\label{sect:dis2}
Previous works \citep[e.g.,][]{Macleod2010, Macleod2012, Kozlowshi2016} aimed to find correlation 
between quasar variability as a function of $L_{\rm bol}$ and $M_{\rm BH}$. Often in these works, 
$M_{\rm BH}$ is estimated via the single-epoch virial black hole mass estimators, i.e., 
$\log M_{\mathrm{BH}}=p_0+p_1\log L+p_2\log\mathrm{FWHM}$, where $p_0$, $p_1$ and $p_3$ 
are constants 
\citep[e.g.,][]{Vestergaard2002, Vestergaard2006, Shen2011}. However, as we demonstrated in 
Section~\ref{sect:sf_rfe} and Figures~\ref{fig:sf_fwhm} \& \ref{fig:sf_rfe_fwhm}, there is no 
clear relation between quasar variability and FWHM. Therefore, we relate quasar variability 
to $L_{\rm bol}$ and \Rfe. 

The main purpose of this section is to provide new empirical relations for future variability 
modeling. Therefore, for simplicity, we assume quasar variability is a CAR(1) process \citep[which 
can, in practice, discribe the light curves well; see, e.g.,][]{Kelly2009, Macleod2010, Macleod2012, 
Zu2013}. 

We aim to explore the correlations between the CAR(1) parameters (i.e., $\tau$ and $\hat{\sigma}$) 
and quasar properties (i.e., $L_{\mathrm{bol}}$ and \Rfe). Following \cite{Kozlowshi2016}, we 
constrain $\hat{\sigma}$ and $\tau$ by modeling the ensemble structure function with 
\begin{equation}
\label{eq:sfmodel}
\mathrm{SF}^2(\Delta t|\tau, \hat{\sigma})=\hat{\sigma}^2\tau(1-\exp(-\Delta t/\tau)^{\beta}) 
+ \sigma_{\mathrm{p}}^2 \\, 
\end{equation}
where $\sigma_{\mathrm{p}}$ is the uncertainty of the magnitude difference between two observations  
separated by $\Delta t$. We fix $\beta=1$ (i.e., the CAR(1) process, see Eq.~\ref{eq:sfdrw}) 
in our subsequent analysis (we will try to set $\beta$ as a free parameter in Section~\ref{sect:dis3}). 

\begin{deluxetable}{ccccc}
\tablecaption{Priors of the parameters \label{table:prior}}
\tablehead{\colhead{} & \colhead{Parameter} & \colhead{Min} & \colhead{Max} & \colhead{Distribution} } 
\startdata
{} & $c_{1i}$ & $0.0$ & $4.0$ & Uniform \\
{Eqs.~\ref{eq:lk1} \& \ref{eq:lk2}} ($i=1$) & $c_{2i}$ & $-10.0$ & $10.0$ & Uniform \\
{or} & $\kappa_{1i}$ & $-2.0$ & $2.0$ & Uniform \\
{Eqs.~\ref{eq:rf1} \& \ref{eq:rf2}} ($i=2$) & $\kappa_{2i}$ & $-2.0$ & $2.0$ & Uniform \\
{} & $\ln \sigma_{\rm int}$ & $-10.0$ & $10.0$ & Uniform \\
\hline
{} & $a$ & $-20.0$ & $6.0$ & Uniform \\
{Eq.~\ref{eq:fitfun}} & $b$ & $-5.0$ & $5.0$ & Uniform \\
{} & $c_1$ & $-5.0$ & $5.0$ & Uniform \\
{} & $\ln \sigma_{\rm int}$ & $-10.0$ & $10.0$ & Uniform \\
\hline
{} & $a$ & $-20.0$ & $6.0$ & Uniform \\
{} & $b$ & $-5.0$ & $5.0$ & Uniform \\
{Eqs.~\ref{eq:cosmo}} & $\Omega_{\mathrm{m}}$ & $0.0$ & $1.0$ & Uniform \\
{} & $\Omega_{\Lambda}$ & $0.0$ & $1.0$ & Uniform \\
{} & $\ln \sigma_{\rm int}$ & $-10.0$ & $10.0$ & Uniform \\
\enddata
\end{deluxetable} 

\begin{figure*}
\epsscale{1.0}
\plotone{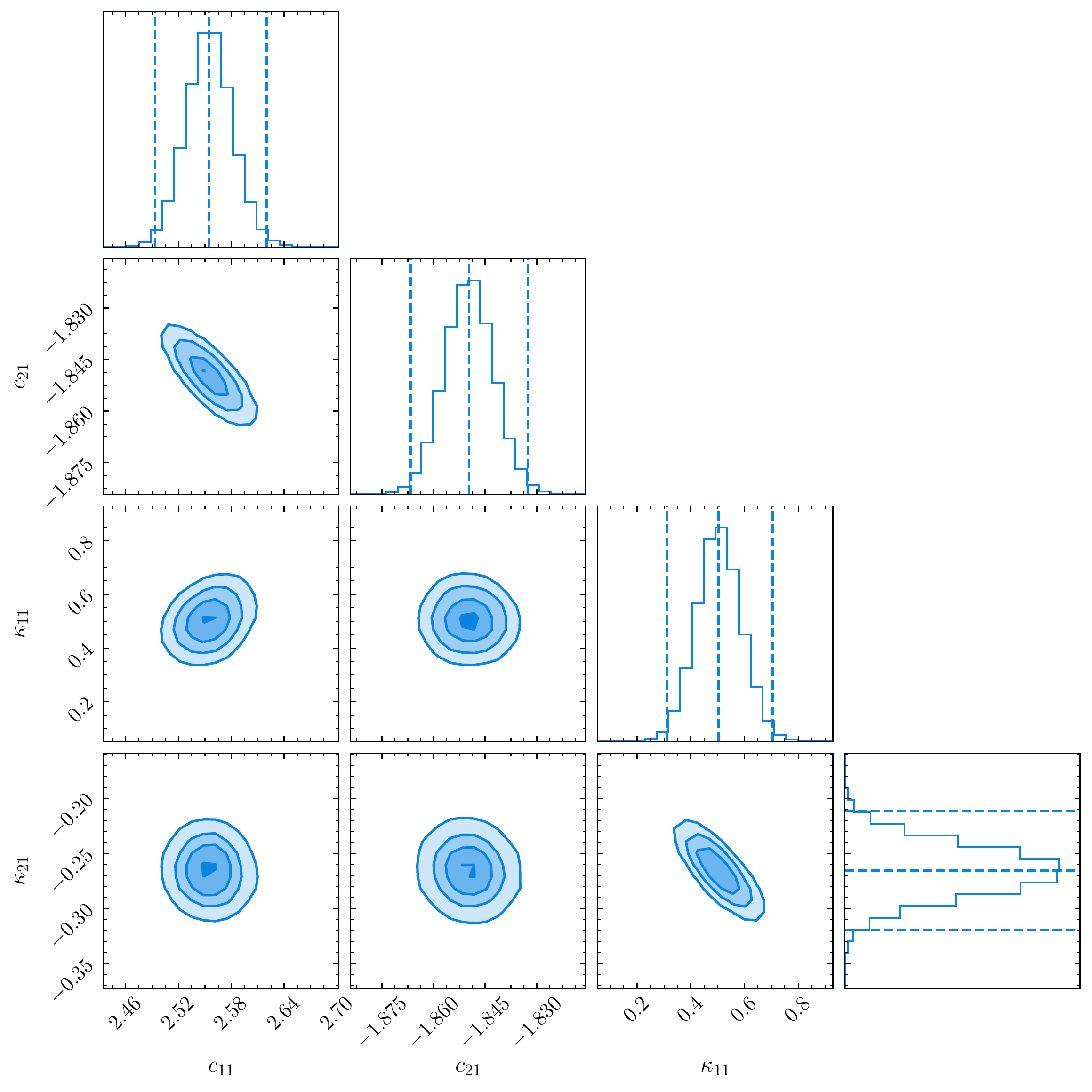}
\caption{The posterior distributions of the parameters for the ensemble structure function 
as a function of $L_{\mathrm{bol}}$. For this figure and subsequent figures, the dashed 
lines indicate the $1$-th, $50$-th, and $99$-th percentiles. The contours indicate the joint 
distributions of two parameters. }
\label{fig:link_l}
\end{figure*}

To explore the dependence of $\tau$ on $L_{\mathrm{bol}}$ and \Rfe, we perform the following 
analysis. For each bin of the \Rfe-matched sample, we assume the ensemble CAR(1) parameters 
are determined by 
\begin{equation}
\label{eq:lk1}
\log \tau = c_{11} + \kappa_{11} \log(\bar{L}_{\mathrm{bol}}/10^{45.5}\ \mathrm{erg\ s^{-1}})\\,
\end{equation}
and 
\begin{equation}
\label{eq:lk2}
\log \hat{\sigma} = c_{21} + \kappa_{21} \log(\bar{L}_{\mathrm{bol}}/10^{45.5}\ \mathrm{erg\ 
s^{-1}})\\,
\end{equation}
where $\bar{L}_{\mathrm{bol}}$ is the average of $L_{\mathrm{bol}}$ in each bin. We also try 
to remove galaxy contamination to $\bar{L}_{\mathrm{bol}}$ by applying the empirical relation 
of Eq.~(1) in \cite{Shen2011}. We then calculate the theoretical structure function from 
these two equations and Eq.~\ref{eq:sfmodel}. 

We fit the theoretical structure functions to the three ensemble structure functions of the 
\Rfe-matched sample via a Bayesian approach. The \textit{likelihood function} is 
\begin{equation}
\label{eq:like2}
\begin{split}
& \ln p(f|x, \mathrm{pms}, \sigma_{\mathrm{int}}) = \\& -\frac{1}{2}\sum_{i=1}^{i=3} 
\sum_{n}\left[\frac{(f_{n, i}-f_{\mathrm{model}, n, i})^2}{s_{n, i}^2} + \ln(2\pi s_{n, i}^2)\right] ,
\end{split}
\end{equation}
where $x$ represents a set of quasar parameters (e.g., $L_{\mathrm{bol}}$, \Rfe); 
$\mathrm{pms}$ is a collection of parameters $c_{11}$, $c_{21}$, $\kappa_{11}$ \& $\kappa_{21}$; 
$f_{\mathrm{model, n, i}}$ and $f_{n, i}$ are the theoretical and observational structure functions, 
respectively; $i=1, 2, 3$ represents the three bins; $n$ indicates each $\Delta t(n)$. 
$\sigma_{\mathrm{int}}$ denotes the summation of the 
measurement uncertainty of $f$ and the (possible) intrinsic scatter.\footnote{The 
intrinsic scatter is considered during the fit since the bootstrap method might 
significantly underestimate the errors of the ensemble structure functions 
\citep{Emmanoulopoulos2010}.} That is, $s_{n, i}^2= (\sigma_{\mathrm{int}}f_{\mathrm{model}, n, i})^2 
+ f^2_{\mathrm{err}, n, i}$, where $f_{\mathrm{err}, n, i}$ is the bootstrap uncertainty 
of $f_{n, i}$. The priors are summarized in Table~\ref{table:prior}. We use the MCMC 
code, \textit{emcee}, to sample the posterior distributions of the parameters. 

The best-fitting structure functions are the solid lines in Figure~\ref{fig:sf_lbol}. 
The posterior distributions of $c_{11}$, $c_{21}$, $\kappa_{11}$, and $\kappa_{21}$ are shown 
in Figure~\ref{fig:link_l} and are summarized in Table~\ref{table:link_fit}. 

\begin{deluxetable}{ccc}
\tablecaption{Statistical properties of the parameters for the ensemble structure function 
as a function of $L_{\mathrm{bol}}$ or \Rfe 
\label{table:link_fit}}
\tablehead{\colhead{} & \colhead{Parameter} & \colhead{Median $\pm$ NIQR} } 
\startdata
{} & $c_{11}$ & $2.55\pm 0.03$ \\
{} & $c_{21}$ & $-1.85\pm 0.01$ \\
{Eqs.~\ref{eq:lk1} \& \ref{eq:lk2}} & $\kappa_{11}$ & $0.50\pm 0.08$ \\
{} & $\kappa_{21}$ & $-0.26\pm 0.02$ \\
{} & $\ln \sigma_{\rm int}$ & $-2.95\pm 0.10$ \\
\hline
\hline
{} & $c_{12}$ & $2.38\pm 0.07$ \\
{} & $c_{22}$ & $-1.72\pm 0.02$ \\
{Eqs.~\ref{eq:rf1} \& \ref{eq:rf2}} & $\kappa_{12}$ & $0.001\pm 0.070$ \\
{} & $\kappa_{22}$ & $-0.08\pm 0.02$ \\
{} & $\ln \sigma_{\rm int}$ & $-2.41\pm 0.10$ \\
\hline
{} & $c_{12}$ & $2.38\pm 0.04$ \\
{} & $c_{22}$ & $-1.72\pm 0.02$ \\
{Eqs.~\ref{eq:rf1} \& \ref{eq:rf2} (with $\kappa_{12} \equiv 0$)} & $\kappa_{12}$ 
& $0$ (fixed) \\
{} & $\kappa_{22}$ & $-0.08\pm 0.01$ \\
{} & $\ln \sigma_{\rm int}$ & $-2.42\pm 0.10$ \\
\enddata
\end{deluxetable}

The correlation (i.e., the slope $\kappa_{11}$) between $\tau$ and $L_{\rm bol}$ for 
fixed \Rfe\ (or fixed Eddington ratio) might simply reflect the dependence of $\tau$ 
on $M_{\rm BH}$ \citep[e.g.,][]{Kelly2009, Macleod2010, Macleod2012, Kozlowshi2016}. 
If so, we expect a strong correlation between $\tau$ and Eddington ratio (or \Rfe) 
for fixed $L_{\rm bol}$. To test this argument and explore quasar variability as a 
function of \Rfe, we fit the ensemble structure functions of the luminosity-matched 
sample with 
\begin{equation}
\label{eq:rf1}
\log \tau = c_{12} + \kappa_{12} \bar{R}_{\mathrm{Fe\,{II}}}\\,
\end{equation}
and 
\begin{equation}
\label{eq:rf2}
\log \hat{\sigma} = c_{22} + \kappa_{22} \bar{R}_{\mathrm{Fe\,{II}}}\\.
\end{equation}
The priors are summarized in Table~\ref{table:prior}. The statistical properties of the 
distributions are summarized in Table~\ref{table:link_fit}. To our surprise, the correlation 
between $\tau$ and \Rfe\ is statistically insignificant as $\kappa_{12}= 
0.001\pm 0.070$. Therefore, we conclude that $\tau$ depends mostly on $L_{\rm bol}$. 

We then refit the ensemble structure functions of the luminosity-matched sample with 
Eq.~\ref{eq:rf1} \& \ref{eq:rf2} but fix $\kappa_{12} \equiv 0$ (i.e., we assume $\tau$ 
does not depend on \Rfe). The statistical properties of the distributions are summarized 
in Table~\ref{table:link_fit}. The best-fitting structure functions are the solid lines 
in Figure~\ref{fig:sf_rfe}. By fixing $\kappa_{12} \equiv 0$, the intrinsic scatter of 
the fit ($\ln \sigma_{\rm int}=-2.41$) is similar to that of the 
previous fit ($\ln \sigma_{\rm int}=-2.42$). That is, $\tau$ and \Rfe\ 
are not tightly correlated. 

Combining the best-fitting relations for the \Rfe- and luminosity-matched samples, we can 
derive quasar variability as a function of $L_{\rm bol}$ and \Rfe, i.e., 
\begin{equation}
\label{eq:md1}
\log \tau = 2.49 + 0.50(\log L_{\mathrm{bol}}-45.50)\\,
\end{equation}
and
\begin{equation}
\label{eq:md2}
\log \hat{\sigma} = -1.788 - 0.26(\log L_{\mathrm{bol}}-45.50) - 0.08
R_{\mathrm{Fe\,{\sc II}}} \\.
\end{equation}

For each S82 quasar with ``good'' data (e.g., at least ten epochs and small measurement 
errors), \cite{Macleod2010} fit the CAR(1) process to 
the light curve and constrained $\hat{\sigma}$ and $\tau$. In principle, we can adopt their 
data and fit the best-fitting parameters as a function of quasar properties. However, 
\cite{Kozlowshi2017} recently demonstrated that, if the baseline is not $\sim 5$--$10$ times 
larger than $\tau$, the best-fitting CAR(1) parameters are biased. The biases are negligible 
for $\hat{\sigma}$ but are rather strong for $\tau$. Therefore, we should only focus on 
$\hat{\sigma}$. 

The function we use to relate $\hat{\sigma}$, $L_{\rm bol}$ and \Rfe\ is:
\begin{equation}
\label{eq:fitfun}
\log \hat{\sigma} = a + b (\log L_{\mathrm{bol}}-45) + 
c_1 R_{\mathrm{Fe\,{II}}} \\.
\end{equation}
For comparison, we also try to fit the following function: 
\begin{equation}
\label{eq:fitfun2}
\log \hat{\sigma} = a + b (\log L_{\mathrm{bol}}-45) + 
c_2 \log \mathrm{FWHM} \\.
\end{equation}

We fit the functions Eq.~\ref{eq:fitfun} \& \ref{eq:fitfun2} to $0.5<z<0.89$ (i.e., a narrow 
range of redshift) quasars via a Bayesian approach. The \textit{likelihood function} is 
\begin{equation}
\label{eq:like}
\begin{split}
&\ln p(\log \hat{\sigma}|x, \sigma_{x}, \mathrm{pms}, \sigma_{\mathrm{int}})= \\& -\frac{1}{2}\sum_{n} 
\left[\frac{(\log \hat{\sigma}_{n}-\log \hat{\sigma}_{\mathrm{model, n}})^2}{s_n^2}+\ln(2\pi s_n^2)\right] ,
\end{split}
\end{equation}
where $x$ is $[\log(L_{\mathrm{bol}}), R_{\mathrm{Fe\,{II}}}]$ (or [$\log(L_{\mathrm{bol}}), 
\mathrm{FWHM}$]); $\sigma_x$ is the uncertainty of $x$; $\mathrm{pms}$ represents parameters 
$a$, $b$ and $c$; $\hat{\sigma}_{\mathrm{model}}$ is given by Eq.~\ref{eq:fitfun} or 
\ref{eq:fitfun2}; $\sigma_{\mathrm{int}}$ is a summation of the measurement uncertainty of 
$\hat{\sigma}$ and the intrinsic scatter. $s_n^2= \sigma_{\mathrm{int}}^2+(b\sigma_{L})^2+
(c\sigma_{c})^2$, where $\sigma_{L}$ and $\sigma_{c}$ are the measurement errors of $L_{\rm bol}$ 
and \Rfe\ (or FWHM), respectively. $\sigma_{\mathrm{int}}$ represents the statistical dispersion 
due to either measurement errors of $\log \hat{\sigma}$ or the intrinsic scatter. The priors are 
summarized in Table~\ref{table:prior}.   

The statistical properties of the parameters $a$, $b$, $c_1$ and $\sigma_{\mathrm{int}}$ for $\hat{\sigma}$ 
as a function of $L_{\rm bol}$ and \Rfe\ (i.e., Eq.~\ref{eq:fitfun}) are presented in 
Table~\ref{table:fitfun}. Our results indicate that while the short-term variability 
is mainly driven by $L_{\rm bol}$, an additional dependence on \Rfe\ (or Eddington ratio) is 
also statistically significant. 

\begin{deluxetable}{ccc}
\tablecaption{Statistical properties of the parameters for the CAR(1) parameter $\hat{\sigma}$ as 
a function of quasar properties \label{table:fitfun}}
\tablehead{\colhead{} & \colhead{Parameter} & \colhead{Median $\pm$ NIQR} } 
\startdata
{}& $a$ & $-1.70\pm 0.02$ \\
{Eq.~\ref{eq:fitfun}}& $b$ & $-0.29\pm 0.02$ \\
{}& $c_1$ & $-0.05\pm 0.01$ \\
{}& $\ln \sigma_{\rm int}$ & $-1.81\pm 0.03$ \\
\hline
{}&$a$ & $-1.86\pm 0.11$ \\
{Eq.~\ref{eq:fitfun2}}&$b$ & $-0.30\pm 0.02$ \\
{}&$c_2$ & $-0.03\pm 0.02$ \\
{}&$\ln \sigma_{\rm int}$ & $-1.78\pm 0.02$ \\
\enddata
\end{deluxetable}

In the works of \cite{Macleod2010} and \cite{Kozlowshi2016}, they explored the 
dependencies of $\mathrm{SF}_{\infty}$ and $\tau$ on $L_{\rm bol}$ and $M_{\rm BH}$. 
Using their best-fitting relations, we can also obtain the relation between 
$\hat{\sigma}$, $L_{\rm bol}$ and Eddington ratio. In both works, the dependence 
of $\hat{\sigma}$ on $L_{\rm bol}$ is close to our result. \cite{Kelly2009, 
Kelly2013} also obtained a similar relation using light curves of the international 
AGN Watch projects.\footnote{For the light curves, please refer to 
\url{http://www.astronomy.ohio-state.edu/~agnwatch/}.} 
However, the correlation between $\hat{\sigma}$ and Eddington ratio is statistically 
insignificant in these works. 

It is quite possible that, in previous works, the correlation between $\hat{\sigma}$ 
and Eddington ratio is diluted by the large uncertainty in $M_{\rm BH}$ due to 
orientation. Indeed, after controlling $L_{\rm bol}$ and Rfe, the ensemble structure 
function does not depend on FWHM (see Figure~\ref{fig:sf_fwhm}). To confirm our guess, 
we explore the dependence of $\hat{\sigma}$ on $L_{\rm bol}$ and FWHM by fitting 
Eq.~\ref{eq:fitfun2}. The priors are summarized in Table~\ref{table:prior}. 
The statistical 
properties of the distributions are summarized in Table~\ref{table:fitfun}. As we 
expected, there is indeed no correlation between $\hat{\sigma}$ and FWHM (the slope, 
$c_2$, is statistically consistent with $0$). Therefore, the additional dependence 
of $\hat{\sigma}$ on Eddington ratio is missed in previous works.

\subsection{Implications to Accretion Physics}
\label{sect:dis3}
In this work, we find the dependence of the variability parameters on $L_{\rm bol}$ 
and \Rfe. Therefore, it is likely that the optical/UV variability is produced in 
the quasar central engine. Several models are proposed to explain the connection 
between the optical/UV variability and quasar properties. For instance, \cite{Li2008} 
proposed that variations in the global accretion rate drive quasar optical/UV 
variability \citep[see also][]{Zuo2012}. However, such model failed to explain 
timescale-dependent color variability \citep[e.g.,][]{SunYH2014, Cai2016, Zhu2016}. 
Instead, models with local fluctuations \citep[possibly regulated by some common variations; 
see][]{Cai2018} in the accretion disk are more compatible with observations. The 
local fluctuation model can also produce the CAR(1) process \citep{Lin2012}. 
Meanwhile, X-ray reprocess might also play a role \citep[e.g.,][]{Czerny1999} 
although no significant correlation between X-ray and UV/optical variations is 
found \citep{Kelly2011, Kelly2013} and the color variability might not be explained 
by X-ray reprocess \citep{Zhu2017}. 

\cite{Kelly2013} proposed that the variance of the short-term variability per 
$\tau_{\rm TH}$ is a constant. If so, for fixed observational timescale, 
$\hat{\sigma}^2 \propto 1/\tau_{\rm TH}$; from the accretion disk theory, we 
expect $\tau_{\rm TH}$ scales with $L_{\mathrm{bol}}^{1/2}$. Therefore, this 
scenario predicts $\hat{\sigma} \propto L_{\mathrm{bol}}^{-1/4}$. This scenario 
can explain our best-fitting relation between $\hat{\sigma}$ and $L_{\rm bol}$ 
(see Tables~\ref{table:link_fit} or \ref{table:fitfun}). 

In contrast to previous works, we find a correlation between $\hat{\sigma}$ and \Rfe. 
The additional dependence of $\hat{\sigma}$ on \Rfe\ might be induced by X-ray 
reprocessing. High-/low-\Rfe\ (Eddington ratio) quasars tend to have weaker/stronger 
X-ray emission \citep{Lusso2012}. As a result, X-ray reprocessing is more efficient 
and can induce more variations in UV/optical bands for low-\Rfe\ sources. A promising 
alternative explanation is that Eddington ratio might correlate with gas metallicity 
\citep[e.g.,][]{Matsuoka2011}. If so, high-\Rfe\ quasars are iron-overabundant, and 
their accretion disks are more stable \citep{Jiang2016}. 

The scatter of $\hat{\sigma}$ as a function of $L_{\rm bol}$ and \Rfe\ is slightly 
smaller than that of the relation between $\hat{\sigma}$, $L_{\rm bol}$ and FWHM. 
These scatters are caused by measurement errors (which is $0.088$ dex) and intrinsic 
scatter. \cite{Guo2017} argued that the intrinsic scatter is caused by the deviation 
from the CAR(1) process on long timescales (see their Figure~9). Based on this spirit, 
they constrained the PSD of quasar variability on long timescales to be steeper than 
$f^{-1.3}$. According to our best-fitting results, the intrinsic scatter in their 
work is slightly over-estimated since they related $\hat{\sigma}$ to FWHM. Therefore, 
the PSD of quasar variability on long timescales approaches the $1/f$ relation. Such 
a PSD is expected from the local variations of accretion rate \citep{Lyubarskii1997, 
Noble2009}. 

We find that the characteristic timescale, $\tau$, is mostly driven by $L_{\rm bol}$ 
(see Section~\ref{sect:dis2}; Table~\ref{table:link_fit}). 
This solo dependence and the normalization encourage us to link $\tau$ with the thermal 
timescale ($\tau_{\mathrm{TH}}$). The best-fitting slope ($0.50\pm 0.08$) is remarkably 
consistent with the theoretical expectation (i.e., the thermal timescale 
$\tau_{\mathrm{TH}}\sim L_{\rm bol}^{0.5}$). It should be noted that, even if $\tau$ 
is the thermal timescale, there might still be an anti-correlation between $\tau$ and 
$M_{\rm BH}$ for fixed $L_{\rm bol}$. This is because the thermal timescale of an 
accretion disk depends positively with iron abundance \citep{Jiang2016}; high Eddington 
ratio quasars might be more metal-rich than low Eddington ratio ones \citep{Matsuoka2011}. 
However, such a correlation is not found in our results. It is possible that this correlation 
is weak and is unable to be revealed in our data. 

Recent works suggested that significant deviations occur on very short timescales \citep[i.e., 
$\sim$ days; see e.g.,][]{Mushotzky2011, Kasliwal2015, Smith2018}. However, on timescales 
we consider here (months to years), this deviation should not be very important. 
\cite{Kozlowshi2016} revealed a positive correlation between $\beta$ and $L_{\rm bol}$ by 
studying the S82 quasars. We then refit 
Eq.~\ref{eq:sfmodel} to the \Rfe -matched samples via the same Bayesian approach 
but set $\beta$ as a free parameter. We do \textit{not} find a significant correlation between 
$\beta$ and $L_{\rm bol}$.  The discrepancy might be caused by the following reasons. First, 
our selected S82 quasars have much lower luminosity than that of \cite{Kozlowshi2016}. 
Second, we use \Rfe\ rather than the ratio of $L_{\rm bol}$ to the virial $M_{\rm BH}$ (which is 
likely biased by orientation) to trace the unknown Eddington ratio. 

The strong correlation between $\tau$ and $L_{\rm bol}$ is also found by \cite{Caplar2017}. 
Note, however, that they adopted a different method to constrain $\tau$. In some other previous 
works \citep{Macleod2010, Kozlowshi2016}, $\tau$ is found to be insensitive to $L_{\rm bol}$ 
but depends on the virial $M_{\rm BH}$. The differences between our results and that of 
\cite{Kozlowshi2016} might also be caused by reasons we mentioned above.\footnote{\cite{Kozlowshi2017} 
argued that $\tau$ can be easily biased toward lower values. The bias anti-correlates with the ratio 
of the (rest-frame) time interval of a light curve to $\tau$. If our $\tau$-$L_{\rm bol}$ relation 
is correct, our best-fitting results are less biased since our selected S82 quasars are 
less luminous (i.e., smaller $\tau$) and have smaller redshifts (i.e., longer rest-frame 
time interval).} 

However, it should be noted that quasar variability on long timescale is likely not 
consistent with the CAR(1) process \citep{Macleod2012, Guo2017}. If so, it is unclear 
that whether we can directly compare $\tau$ with physical timescales. The forthcoming 
era of time domain astronomy is the key to answer the physical nature of $\tau$.

\subsection{Quasar Variability as a probe of Cosmology?}
\label{sect:dis4}
Our work and many previous works \citep[e.g.,][]{Macleod2010, Kozlowshi2016, Caplar2017} 
indicate that the short-term UV/optical variability amplitude (or $\hat{\sigma}$) depends 
critically on $L_{\rm bol}$. In this work, we also find an additional dependence of 
$\hat{\sigma}$ on \Rfe. This additional dependence is statistically significant but rather 
weak since the slope is $-0.05\pm 0.01$ (see Table~\ref{table:fitfun}). In practice, 
we can ignore this additional dependence and fit $\hat{\sigma}$ only as a function of 
$L_{\rm bol}$ (i.e., the parameter $c_1$ in Eq.~\ref{eq:fitfun} is fixed to be $0$). 
The best-fitting parameters are $\hat{\sigma}=(-1.74\pm 0.012)-(0.30\pm 0.018) 
L_{\mathrm{bol}}$; the scatter (i.e., $\sigma_{\mathrm{int}}$ which is a combination of 
measurement errors and the intrinsic scatter) is $\exp(-1.81\pm0.026)$ which is the same 
as that of Eq.~\ref{eq:fitfun}. We can, in principle, estimate $L_{\rm bol}$ from the 
$\hat{\sigma}$-$L_{\rm bol}$ relation without assuming any cosmological models. Therefore, 
it is possible to use quasar short-term UV/optical variability as a probe of cosmology 
parameters. 

The $\hat{\sigma}$-$L_{\rm bol}$ relation (i.e., Eq.~\ref{eq:fitfun}  with $c_1$ is fixed 
to be zero) can be revised as 
\begin{equation}
\label{eq:cosmo}
\log \hat{\sigma} = a + b\log(4\pi f/10^{45}\ \mathrm{erg\ s^{-1}}) + 2b\log(D_{L}) \\, 
\end{equation}
where $f$ is the observed flux. $D_L$, the luminosity distance, is a function of cosmological 
model and can be independently measured if we know $\hat{\sigma}$, $f$, $a$ and $b$. Given 
the intrinsic scatter of the $\hat{\sigma}$-$L_{\rm bol}$ relation, such constraints can 
be made only with a large sample of quasars that span over a wide range of cosmic history. 

\begin{figure}
\epsscale{1.2}
\plotone{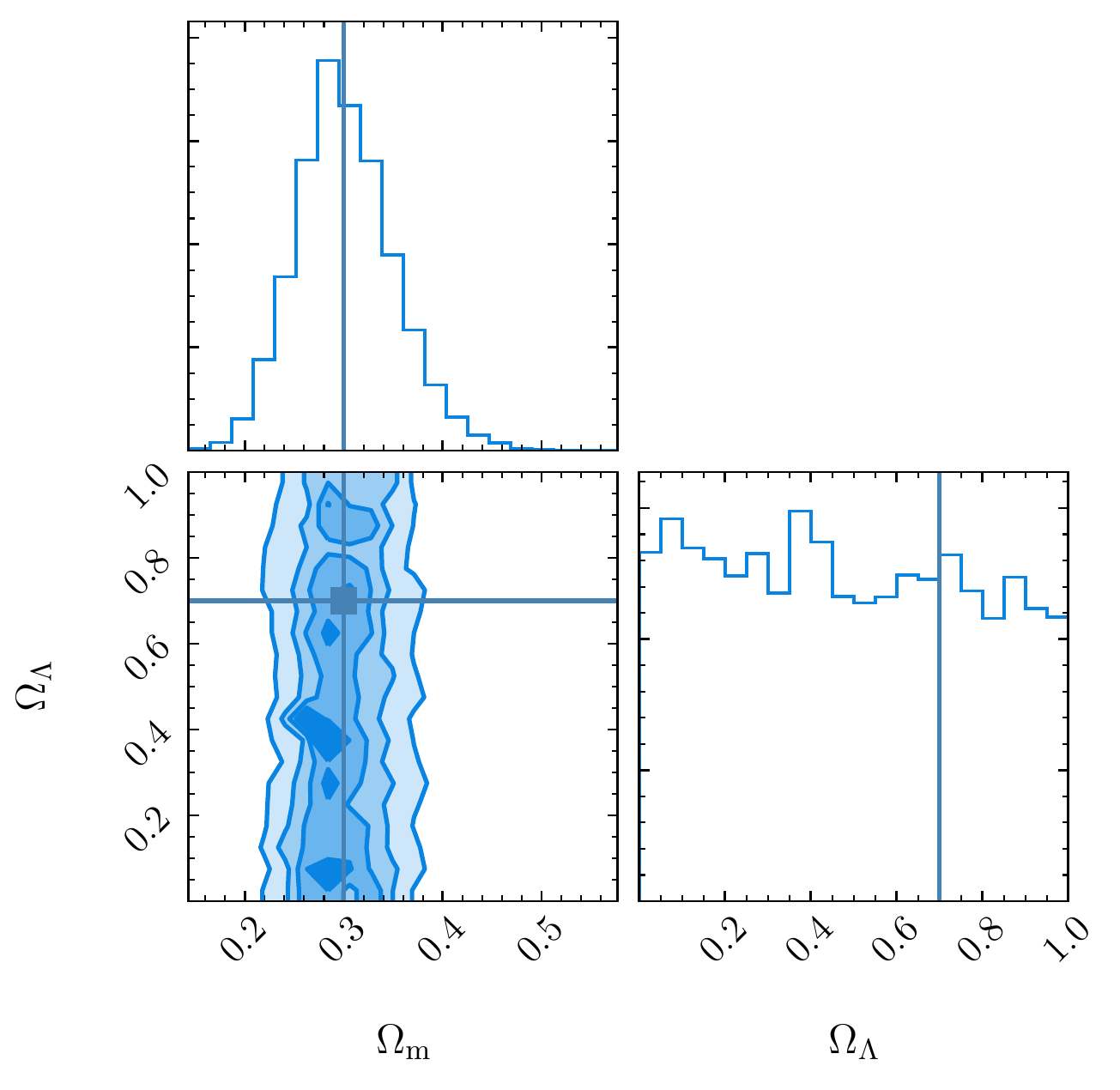}
\caption{The posterior distributions of the parameters for $\hat{\sigma}$ as a function 
of $L_{\mathrm{bol}}$ and cosmology parameters ($\Omega_{\rm m}$ and $\Omega_{\lambda}$) 
from $10^5$ simulated quasars. The blue lines and dot indicate the input parameters. 
$\Omega_{\rm m}$ is well constrained to be $0.28\pm 0.03$. On the other hand, 
$\Omega_{\lambda}$ cannot be constrained. }
\label{fig:cosmo_simu}
\end{figure}

To illustrate this idea, we perform the following simulation of $10^5$ quasars. For each 
quasar, the intrinsic $L_{\rm bol}$ and \Rfe\ and their measurement errors are assigned 
according to the randomly selected quasar from our parent sample. We then calculate 
$\hat{\sigma}_{\rm th}$ from our best-fitting Eq.~\ref{eq:fitfun}; a Gaussian noise with 
standard deviation of $\exp(-1.81)$ (see 
Section~\ref{sect:dis2}) is added to $\hat{\sigma}_{\rm th}$ to generate the observed 
$\hat{\sigma}$. We also assign galaxy contamination according to Eq.(1) of \cite{Shen2011}. 
The observed $\hat{\sigma}$ is diluted by the non-variable galaxy emission. The observed 
$L_{\rm bol}$ and \Rfe\ are generated by perturbing intrinsic $L_{\rm bol}$ and \Rfe\ 
with their measurement errors. In addition, the galaxy emission is added to the observed 
$L_{\rm bol}$. To calculate the observed flux, we assume a flat $\Lambda$CDM cosmology 
with $h_0=0.7$ and $\Omega_{\mathrm{m}}=0.3$; redshift is randomly assigned from a uniform 
distribution within [$0.1$, $0.89$]. We then fit Eq.~\ref{eq:cosmo} to the simulated mock 
sample by considering the $\Lambda$CDM cosmology with $h_0=0.7$ via a Bayesian approach. 
Both $\Omega_{\mathrm{m}}$ (i.e., the matter density fraction) and $\Omega_{\Lambda}$ 
(i.e., the dark energy fraction) are free parameters. The \textit{likelihood function} 
is the same as Eq.~\ref{eq:like} and the priors are summarized in Table~\ref{table:prior}.  

The posterior distributions of the model parameters are presented in 
Figure~\ref{fig:cosmo_simu}. Even if $\Omega_{\Lambda}$ is not constrained, the recovered 
$\Omega_{\mathrm{m}}=0.28\pm 0.03$ is accurate. Note that, if the sample size is limited to 
$10^4$, the recovered $\Omega_{\mathrm{m}}=0.36\pm 0.12$. Therefore, the large sample size 
is one of the key factor. 

Our simulated sample might be available in the era of time-domain astronomy (e.g., with 
the Large Synoptic Survey Telescope). However, it remains unclear whether the scatter 
of the $\hat{\sigma}$-$L_{\rm bol}$ relation depends on the sample size/redshift 
or not. In order 
to test this hypothesis, we select sources with $0.96<z<1.48$ or $1.48<z<2.03$. Their 
$r$ or $i$ bands correspond to the rest-frame of $0.5<z<0.9$ $g$ band. We calculate 
the differences between their $\hat{\sigma}$ and the expectation values from our best-fitting 
$\hat{\sigma}$-$L_{\rm bol}$ relation. Some of the differences are due to a combination 
of the scatter of the relation and the measurement error of $L_{\rm bol}$, $\sigma_{L}$ 
(i.e., the total scatter is $\sqrt{\sigma_{\mathrm{int}}^2+(b\sigma_{L})^2}$). We then 
calculate the ratio of the differences to this total scatter. We find that, for $95\%$ 
of sources, the ratio is less than $3$. For the remaining $5\%$ sources, many of them are 
highly variable ones ($20\%$ have $\hat{\sigma}>1$). Such sources might be ``changing-look'' 
AGN candidates \citep{Macleod2016}; the origin of such variability could be different. 
Therefore, it is unlikely that the scatter of the $\hat{\sigma}$-$L_{\rm bol}$ relation 
significantly depends on the sample size/redshift. We can also 
use high-redshift ($z\sim 2$) quasars to constrain cosmological parameters; the accuracy 
would be further improved. 

In addition to our method, it is also proposed that the BLR and dust reverberation 
\citep{Watson2011, Yoshii2014}, the nonlinear relation between the ultraviolet and 
X-ray luminosities \citep{Risaliti2015}, the X-ray variability and broad line width 
\citep{Franca2014}, and the saturated luminosity of super-Eddington AGNs 
\citep{Wang2013} can also be adopted as distance measurements. In conclusion, AGNs 
will play a more important role in measuring the Universe \citep[for a recent review, 
see][]{Czerny2018}.

\section{Summary}
\label{sect:summ}
In this work, we have explored the evolution of the optical g-band variability 
of SDSS S82 quasars along the quasar main sequence. Our study focuses 
on quasar variability on timescales of weeks to years. Our main results are as 
follows. 
\begin{itemize}
\item[1.] The variability amplitude decreases with $L_{\rm bol}$ (Section~\ref{sect:sf_lbol}; 
Figure~\ref{fig:sf_lbol}) and \Rfe\ (Section~\ref{sect:sf_rfe}; Figure~\ref{fig:sf_rfe}). 
After controlling luminosity and \Rfe, high- and low-FWHM sources show similar variability 
(Figure~\ref{fig:sf_fwhm}). These results support the scenario that \Rfe\ is governed by 
Eddington ratio \citep{Shen2014}; FWHM traces orientation (Section~\ref{sect:dis1}). 

\item[2.] We provide new empirical relations between variability parameters, $L_{\rm bol}$ 
and \Rfe\ (Section~\ref{sect:dis2}; Eq.~\ref{eq:md1} \& \ref{eq:md2}). 

\item[3.] Our new empirical relations are consistent with the scenario that quasar variability 
is driven by the thermal fluctuations in the accretion disk; $\tau$ seems to correspond to the 
thermal timescale (Section~\ref{sect:dis3}). X-ray reprocessing and/or gas metallicity might 
also play a role in determining short-term variability. 

\item[4.] The short-term variability depends mostly upon $L_{\rm bol}$. We then propose that 
short-term (a few months) quasar variability might be regarded as a new type of ``Standard Candle''. Our 
simple simulation suggests that the cosmological parameters can be well constrained with a 
sample of $10^5$ quasars (Section~\ref{sect:dis4}). 
\end{itemize}
In this work, we only focus on the SDSS S82 quasars. Therefore, we cannot constrain 
quasar variability on timescales of sub-months. On such timescales, it has been shown that 
the PSD of quasar variability has an additional break to $f^{-n}$ with $n>2$ \citep{Mushotzky2011, 
Kasliwal2015}. It would also be interesting to explore the relation between such variability 
and \Rfe. Meanwhile, current and future surveys, e.g., SDSS, PTF \citep{Law2009}, DES 
\citep{Honscheid2008} and LSST \citep{Ivezic2008} can provide much better light curves in 
terms of cadence and baseline. Our results can be justified and extrapolated in the era of 
time domain astronomy.

\acknowledgments
We thank the anonymous scientific and statistical referees for their helpful comments that improved 
the paper. M.Y.S., Y.Q.X., J.X.W. and Z.Y.C. acknowledge the support from NSFC-11603022, NSFC-11473026, 
NSFC-11421303, NSFC-11503024, the 973 Program (2015CB857004, 2015CB857005), the China 
Postdoctoral Science Foundation (2016M600485), the CAS Frontier Science Key Research Program 
(QYZDJ-SSW-SLH006). 

Funding for the SDSS and SDSS-II has been provided by the Alfred P. Sloan Foundation, the 
Participating Institutions, the National Science Foundation, the U.S. Department of Energy, 
the National Aeronautics and Space Administration, the Japanese Monbukagakusho, the Max 
Planck Society, and the Higher Education Funding Council for England. The SDSS Web site is 
http://www.sdss.org/.

The SDSS is managed by the Astrophysical Research Consortium for the Participating Institutions. 
The Participating Institutions are the American Museum of Natural History, Astrophysical 
Institute Potsdam, University of Basel, Univer- sity of Cambridge, Case Western Reserve 
University, University of Chicago, Drexel University, Fermilab, the Institute for Advanced 
Study, the Japan Participation Group, Johns Hopkins University, the Joint Institute for 
Nuclear Astrophysics, the Kavli Institute for Particle Astrophysics and Cosmology, the 
Korean Scientist Group, the Chinese Academy of Sciences (LAMOST), Los Alamos National 
Laboratory, the Max-Planck Institute for Astronomy (MPIA), the Max-Planck-Institute for 
Astrophysics (MPA), New Mexico State University, Ohio State University, University of 
Pittsburgh, University of Portsmouth, Princeton University, the United States Naval 
Observatory, and the University of Washington.

This publication makes use of data products from the Wide-field Infrared Survey Explorer, 
which is a joint project of the University of California, Los Angeles, and the Jet Propulsion 
Laboratory/California Institute of Technology, and NEOWISE, which is a project of the Jet 
Propulsion Laboratory/California Institute of Technology. WISE and NEOWISE are funded by 
the National Aeronautics and Space Administration.

\software{Astropy \citep{Astropy2013}, CARMA \citep{Kelly2014}, Matplotlib \citep{Hunter2007}, 
Numpy \& Scipy \citep{scipy}}, \textit{emcee} \citep{emcee}, \textit{corner.py} \citep{corner}.


\begin{thebibliography}{}
\bibitem[Ai et al.(2010)]{Ai2010} Ai, Y.~L., Yuan, W., Zhou, H.~Y., et al.\ 2010, \apjl, 716, L31
\bibitem[Akaike(1974)]{Akaike1974} Akaike, H.\ 1974, IEEE Transactions on Automatic Control, 19, 716
\bibitem[Astropy Collaboration et al.(2013)]{Astropy2013} Astropy Collaboration, Robitaille, T.~P., Tollerud, E.~J., et al.\ 2013, \aap, 558, A33 
\bibitem[Bauer et al.(2009)]{Bauer2009} Bauer, A., Baltay, C., Coppi, P., Ellman, N., Jerke, J., Rabinowitz, D. \& Scalzo, R. 2009, ApJ, 696, 1241
\bibitem[Bentz et al.(2013)]{Bentz2013} Bentz, M.~C., Denney, K.~D., Grier, C.~J., et al.\ 2013, \apj, 767, 149
\bibitem[Bisogni et al.(2017)]{Bisogni2017} Bisogni, S., Marconi, A., \& Risaliti, G.\ 2017, \mnras, 464, 385
\bibitem[Boroson(2002)]{Boroson2002} Boroson, T.~A.\ 2002, \apj, 565, 78
\bibitem[Boroson \& Green(1992)]{Boroson1992} Boroson, T.~A., \& Green, R.~F.\ 1992, \apjs, 80, 109 
\bibitem[Cackett et al.(2017)]{Cackett2017} Cackett, E.~M., Chiang, C.-Y., McHardy, I., et al.\ 2018, \apj, 857, 53 
\bibitem[Cai et al.(2016)]{Cai2016} Cai, Z.-Y., Wang, J.-X., Gu, W.-M., et al.\ 2016, \apj, 826, 7
\bibitem[Cai et al.(2018)]{Cai2018} Cai, Z.-Y., Wang, J.-X., Zhu, F.-F., et al.\ 2018, \apj, 855, 117
\bibitem[Caplar et al.(2017)]{Caplar2017} Caplar, N., Lilly, S.~J., \& Trakhtenbrot, B.\ 2017, \apj, 834, 111 
\bibitem[Collier et al.(1998)]{Collier1998} Collier, S.~J., Horne, K., Kaspi, S., et al.\ 1998, \apj, 500, 162
\bibitem[Collin et al.(2006)]{Collin2006} Collin, S., Kawaguchi, T., Peterson, B.~M., \& Vestergaard, M.\ 2006, \aap, 456, 75
\bibitem[Czerny et al.(2018)]{Czerny2018} Czerny, B., Beaton, R., Bejger, M., et al.\ 2018, \ssr, 214, \#32
\bibitem[Czerny et al.(1999)]{Czerny1999} Czerny, B., Schwarzenberg-Czerny, A., \& Loska, Z.\ 1999, \mnras, 303, 148
\bibitem[de Vries et al.(2005)]{Vries2005} de Vries, W. H., Becker, R. H., White, R. L. \& Loomis, C. 2005, AJ, 129, 615
\bibitem[Du et al.(2014)]{Du2014} Du, P., Hu, C., Lu, K.-X., et al.\ 2014, \apj, 782, 45
\bibitem[Edelson et al.(1996)]{Edelson1996} Edelson, R.~A., Alexander, T., Crenshaw, D.~M., et al.\ 1996, \apj, 470, 364
\bibitem[Edelson et al.(2017)]{Edelson2017} Edelson, R., Gelbord, J., Cackett, E., et al.\ 2017, \apj, 840, 41 
\bibitem[Edelson et al.(2015)]{Edelson2015} Edelson, R., Gelbord, J.~M., Horne, K., et al.\ 2015, \apj, 806, 129 
\bibitem[Emmanoulopoulos et al.(2010)]{Emmanoulopoulos2010} Emmanoulopoulos, D., McHardy, I.~M., \& Uttley, P.\ 2010, \mnras, 404, 931
\bibitem[Fausnaugh et al.(2016)]{Fausnaugh2016} Fausnaugh, M.~M., Denney, K.~D., Barth, A.~J., et al.\ 2016, \apj, 821, 56
\bibitem[Foreman-Mackey(2016)]{corner} Foreman-Mackey, D.\ 2016, The Journal of Open Source Software, 2016,
\bibitem[Foreman-Mackey et al.(2013)]{emcee} Foreman-Mackey, D., Hogg, D.~W., Lang, D., \& Goodman, J.\ 2013, \pasp, 125, 306 
\bibitem[Gardner \& Done(2017)]{Gardner2017} Gardner, E., \& Done, C.\ 2017, \mnras, 470, 3591
\bibitem[Giveon et al.(1999)]{Giveon1999} Giveon, U., Maoz, D., Kaspi, S., Netzer, H. \& Smith, P.  S. 1999, MNRAS, 306, 637
\bibitem[Grier et al.(2017a)]{Grier2017a} Grier, C.~J., Pancoast, A., Barth, A.~J., et al.\ 2017a, \apj, 849, 146
\bibitem[Grier et al.(2017b)]{Grier2017b} Grier, C.~J., Trump, J.~R., Shen, Y., et al.\ 2017b, \apj, 851, 21
\bibitem[Gunn et al.(2006)]{Gunn2006} Gunn, J.~E., Siegmund, W.~A., Mannery, E.~J., et al.\ 2006, \aj, 131, 2332
\bibitem[Guo et al.(2017)]{Guo2017} Guo, H., Wang, J., Cai, Z., \& Sun, M.\ 2017, \apj, 847, 132 
\bibitem[Hawkins(2002)]{Hawkins2002} Hawkins, M. R. S. 2002, MNRAS, 329, 76
\bibitem[Honscheid et al.(2008)]{Honscheid2008} Honscheid, K., DePoy, D.~L., \& for the DES Collaboration 2008, arXiv:0810.3600
\bibitem[Hook et al.(1994)]{Hook1994} Hook, I. M., McMahon, R. G., Boyle, B. J. \& Irwin, M. J. 1994, MNRAS, 268, 305
\bibitem[Hunter(2007)]{Hunter2007} Hunter, J.~D.\ 2007, Computing in Science and Engineering, 9, 90 
\bibitem[Ivezi{\'c} et al.(2007)]{Ivezic2007} Ivezi{\'c}, {\v Z}., Smith, J.~A., Miknaitis, G., et al.\ 2007, \aj, 134, 973
\bibitem[Ivezic et al.(2008)]{Ivezic2008} Ivezic, Z., Tyson, J.~A., Abel, B., et al.\ 2008, arXiv:0805.2366 
\bibitem[Jarvis \& McLure(2006)]{Jarvis2006} Jarvis, M.~J., \& McLure, R.~J.\ 2006, \mnras, 369, 182
\bibitem[Jiang et al.(2016)]{Jiang2016} Jiang, Y.-F., Davis, S.~W., \& Stone, J.~M.\ 2016, \apj, 827, 10
\bibitem[Kasliwal et al.(2015)]{Kasliwal2015} Kasliwal, V.~P., Vogeley, M.~S., \& Richards, G.~T.\ 2015, \mnras, 451, 4328
\bibitem[Kelly et al.(2009)]{Kelly2009} Kelly, B. C., Bechtold, J. \& Siemiginowska, A. 2009, ApJ, 698, 895
\bibitem[Kelly et al.(2014)]{Kelly2014} Kelly, B.~C., Becker, A.~C., Sobolewska, M., Siemiginowska, A., \& Uttley, P.\ 2014, \apj, 788, 33
\bibitem[Kelly et al.(2011)]{Kelly2011} Kelly, B.~C., Sobolewska, M., \& Siemiginowska, A.\ 2011, \apj, 730, 52
\bibitem[Kelly et al.(2013)]{Kelly2013} Kelly, B.~C., Treu, T., Malkan, M., Pancoast, A., \& Woo, J.-H.\ 2013, \apj, 779, 187 
\bibitem[Koz{\l}owski(2016)]{Kozlowshi2016} Koz{\l}owski, S.\ 2016, \apj, 826, 118 
\bibitem[Koz{\l}owski(2017)]{Kozlowshi2017} Koz{\l}owski, S.\ 2017, \aap, 597, A128
\bibitem[Koz{\l}owski et al.(2010)]{Kozlowski2010} Koz{\l}owski, S., Kochanek, C.~S., Udalski, A., et al.\ 2010, \apj, 708, 927
\bibitem[Krolik et al.(1991)]{Krolik1991} Krolik, J.~H., Horne, K., Kallman, T.~R., et al.\ 1991, \apj, 371, 541
\bibitem[Kubota \& Done(2018)]{Kubota2018} Kubota, A., \& Done, C.\ 2018, arXiv:1804.00171 
\bibitem[La Franca et al.(2014)]{Franca2014} La Franca, F., Bianchi, S., Ponti, G., Branchini, E., \& Matt, G.\ 2014, \apjl, 787, L12 
\bibitem[Law et al.(2009)]{Law2009} Law, N.~M., Kulkarni, S.~R., Dekany, R.~G., et al.\ 2009, \pasp, 121, 1395
\bibitem[Li \& Cao(2008)]{Li2008} Li, S.-L., \& Cao, X.\ 2008, \mnras, 387, L41 
\bibitem[Lin et al.(2012)]{Lin2012} Lin, D.-B., Gu, W.-M., Liu, T., Sun, M.-Y., \& Lu, J.-F.\ 2012, \apj, 761, 29
\bibitem[Lusso et al.(2012)]{Lusso2012} Lusso, E., Comastri, A., Simmons, B.~D., et al.\ 2012, \mnras, 425, 623
\bibitem[Lyubarskii(1997)]{Lyubarskii1997} Lyubarskii, Y.~E.\ 1997, \mnras, 292, 679 
\bibitem[MacLeod et al.(2010)]{Macleod2010} MacLeod, C. L., Ivezi\'{c}, \v{Z}., Kochanek, C. S. et al. 2010, ApJ, 721, 1014
\bibitem[MacLeod et al.(2012)]{Macleod2012} MacLeod, C. L., Ivezi\'{c}, \v{Z}., Sesar, B. et al. 2012, ApJ, 753, 106
\bibitem[MacLeod et al.(2016)]{Macleod2016} MacLeod, C.~L., Ross, N.~P., Lawrence, A., et al.\ 2016, \mnras, 457, 389 
\bibitem[Mainzer et al.(2011)]{Mainzer2011} Mainzer, A., Bauer, J., Grav, T., et al.\ 2011, \apj, 731, 53
\bibitem[Matsuoka et al.(2011)]{Matsuoka2011} Matsuoka, K., Nagao, T., Marconi, A., Maiolino, R., \& Taniguchi, Y.\ 2011, \aap, 527, A100
\bibitem[McHardy et al.(2014)]{McHardy2014} McHardy, I.~M., Cameron, D.~T., Dwelly, T., et al.\ 2014, \mnras, 444, 1469
\bibitem[McHardy et al.(2017)]{McHardy2017} McHardy, I., Connolly, S., Cackett, K.~E., et al.\ 2017, arXiv:1712.04852
\bibitem[McHardy et al.(2016)]{McHardy2016} McHardy, I.~M., Connolly, S.~D., Peterson, B.~M., et al.\ 2016, Astronomische Nachrichten, 337, 500
\bibitem[Mushotzky et al.(2011)]{Mushotzky2011} Mushotzky, R.~F., Edelson, R., Baumgartner, W., \& Gandhi, P.\ 2011, \apjl, 743, L12
\bibitem[Noble \& Krolik(2009)]{Noble2009} Noble, S.~C., \& Krolik, J.~H.\ 2009, \apj, 703, 964
\bibitem[Pancoast et al.(2014)]{Pancoast2014} Pancoast, A., Brewer, B.~J., \& Treu, T.\ 2014, \mnras, 445, 3055
\bibitem[Panda et al.(2017)]{Panda2017} Panda, S., Czerny, B., Wildy, C., \& {\'S}niegowska, M.\ 2017, arXiv:1712.05176
\bibitem[Risaliti \& Lusso(2015)]{Risaliti2015} Risaliti, G., \& Lusso, E.\ 2015, \apj, 815, 33
\bibitem[Risaliti et al.(2011)]{Risaliti2011} Risaliti, G., Salvati, M., \& Marconi, A.\ 2011, \mnras, 411, 2223
\bibitem[Runnoe et al.(2014)]{Runnoe2014} Runnoe, J.~C., Brotherton, M.~S., DiPompeo, M.~A., \& Shang, Z.\ 2014, \mnras, 438, 3263
\bibitem[Runnoe et al.(2013)]{Runnoe2013} Runnoe, J.~C., Brotherton, M.~S., Shang, Z., Wills, B.~J., \& DiPompeo, M.~A.\ 2013, \mnras, 429, 135  
\bibitem[Sergeev et al.(2005)]{Sergeev2005} Sergeev, S.~G., Doroshenko, V.~T., Golubinskiy, Y.~V., Merkulova, N.~I., \& Sergeeva, E.~A.\ 2005, \apj, 622, 129
\bibitem[Sesar et al.(2007)]{Sesar2007} Sesar, B., Ivezi{\'c}, {\v Z}., Lupton, R.~H., et al.\ 2007, \aj, 134, 2236
\bibitem[Shakura \& Sunyaev(1973)]{SSD} Shakura, N.~I., \& Sunyaev, R.~A.\ 1973, \aap, 24, 337
\bibitem[Shappee et al.(2014)]{Shappee2014} Shappee, B.~J., Prieto, J.~L., Grupe, D., et al.\ 2014, \apj, 788, 48
\bibitem[Shen(2013)]{Shen2013} Shen, Y.\ 2013, Bulletin of the Astronomical Society of India, 41, 61
\bibitem[Shen et al.(2015)]{Shen2015} Shen, Y., Brandt, W.~N., Dawson, K.~S., et al.\ 2015, \apjs, 216, 4
\bibitem[Shen \& Ho(2014)]{Shen2014} Shen, Y., \& Ho, L.~C.\ 2014, \nat, 513, 210 
\bibitem[Shen et al.(2011)]{Shen2011} Shen, Y., Richards, G.~T., Strauss, M.~A., et al.\ 2011, \apjs, 194, 45
\bibitem[Simm et al.(2016)]{Simm2016} Simm, T., Salvato, M., Saglia, R., et al.\ 2016, \aap, 585, A129 
\bibitem[Smith et al.(2018)]{Smith2018} Smith, K.~L., Mushotzky, R.~F., Boyd, P.~T., et al.\ 2018, \apj, 857, 141
\bibitem[Sniegowska et al.(2017)]{Sniegowska2017} Sniegowska, M., Czerny, B., You, B., et al.\ 2017, \aap, 613, A38 
\bibitem[Starkey et al.(2017)]{Starkey2017} Starkey, D., Horne, K., Fausnaugh, M.~M., et al.\ 2017, \apj, 835, 65
\bibitem[Starkey et al.(2016)]{Starkey2016} Starkey, D.~A., Horne, K., \& Villforth, C.\ 2016, \mnras, 456, 1960
\bibitem[Storchi-Bergmann et al.(2017)]{Storchi2017} Storchi-Bergmann, T., Schimoia, J.~S., Peterson, B.~M., et al.\ 2017, \apj, 835, 236
\bibitem[Sulentic et al.(2000a)]{Sulentic2000a} Sulentic, J.~W., Marziani, P., \& Dultzin-Hacyan, D.\ 2000a, \araa, 38, 521
\bibitem[Sulentic et al.(2000b)]{Sulentic2000b} Sulentic, J.~W., Zwitter, T., Marziani, P., \& Dultzin-Hacyan, D.\ 2000b, \apjl, 536, L5 
\bibitem[Sun \& Shen(2015)]{SunJY2015} Sun, J., \& Shen, Y.\ 2015, \apjl, 804, L15 
\bibitem[Sun et al.(2015)]{Sun2015} Sun, M., Trump, J.~R., Shen, Y., et al.\ 2015, \apj, 811, 42 
\bibitem[Sun et al.(2014)]{SunYH2014} Sun, Y.-H., Wang, J.-X., Chen, X.-Y., \& Zheng, Z.-Y.\ 2014, \apj, 792, 54
\bibitem[Sun et al.(2018)]{Sun2018} Sun, M., Xue, Y., Cai, Z., \& Guo, H.\ 2018, \apj, 857, 86
\bibitem[Ulrich et al.(1997)]{Ulrich1997} Ulrich, M.-H., Maraschi, L., \& Urry, C.~M.\ 1997, \araa, 35, 445
\bibitem[Uomoto et al.(1976)]{Uomoto1976} Uomoto, A. K., Wills, B. J. \& Wills, D. 1976, AJ, 81, 905
\bibitem[Vanden Berk et al.(2004)]{Berk2004} Vanden Berk, D. E., Wilhite, B. C., Kron, R. G. et al. 2004, ApJ, 601, 692
\bibitem[Van Der Walt et al.(2011)]{scipy} Van Der Walt, S., Colbert, S.~C., \& Varoquaux, G.\ 2011, arXiv:1102.1523 
\bibitem[Vestergaard(2002)]{Vestergaard2002} Vestergaard, M.\ 2002, \apj, 571, 733 
\bibitem[Vestergaard \& Peterson(2006)]{Vestergaard2006} Vestergaard, M., \& Peterson, B.~M.\ 2006, \apj, 641, 689 
\bibitem[Wanders et al.(1997)]{Wanders1997} Wanders, I., Peterson, B.~M., Alloin, D., et al.\ 1997, \apjs, 113, 69
\bibitem[Wang et al.(2013)]{Wang2013} Wang, J.-M., Du, P., Valls-Gabaud, D., Hu, C., \& Netzer, H.\ 2013, \prl, 110, 081301
\bibitem[Watson et al.(2011)]{Watson2011} Watson, D., Denney, K.~D., Vestergaard, M., \& Davis, T.~M.\ 2011, \apjl, 740, L49
\bibitem[White(1982)]{White1982} White, H.\ 1982, Econometrica, 50, 1 
\bibitem[Wilhite et al.(2008)]{Wilhite2008} Wilhite, B.~C., Brunner, R.~J., Grier, C.~J., Schneider, D.~P., \& vanden Berk, D.~E.\ 2008, \mnras, 383, 1232 
\bibitem[Wright et al.(2010)]{Wright2010} Wright, E.~L., Eisenhardt, P.~R.~M., Mainzer, A.~K., et al.\ 2010, \aj, 140, 1868-1881
\bibitem[Xiao et al.(2018)]{Xiao2018} Xiao, M., Du, P., Horne, K.~D., et al.\ 2018, arXiv:1808.00705
\bibitem[Yoshii et al.(2014)]{Yoshii2014} Yoshii, Y., Kobayashi, Y., Minezaki, T., Koshida, S., \& Peterson, B.~A.\ 2014, \apjl, 784, L11
\bibitem[Zhu et al.(2017)]{Zhu2017} Zhu, F.-F., Wang, J.-X., Cai, Z.-Y., et al.\ 2017, \apj, 860, 29
\bibitem[Zhu et al.(2016)]{Zhu2016} Zhu, F.-F., Wang, J.-X., Cai, Z.-Y., \& Sun, Y.-H.\ 2016, \apj, 832, 75
\bibitem[Zu et al.(2013)]{Zu2013} Zu, Y., Kochanek, C.~S., Koz{\l}owski, S., \& Udalski, A.\ 2013, \apj, 765, 106
\bibitem[Zuo et al.(2012)]{Zuo2012} Zuo, W., Wu, X.-B., Liu, Y.-Q., \& Jiao, C.-L.\ 2012, \apj, 758, 104 
\end{thebibliography}
\end{document}